\documentclass{aa}  

\usepackage{graphicx}
\usepackage{txfonts}
\usepackage{hyperref}
\usepackage{gensymb}
\usepackage{newtxtext,newtxmath}
\usepackage{longtable}
\usepackage{setspace}
\usepackage{booktabs}
\usepackage{wallpaper}
\usepackage{epstopdf}
\usepackage{eso-pic}
\usepackage{eqparbox}
\usepackage{multicol}
\usepackage{dcolumn}
\usepackage{listings}
\usepackage{multirow}

\usepackage[singlelinecheck=off]{caption}
\usepackage{longtable}

\usepackage{ae,aecompl}

\usepackage{amsmath}    
\usepackage{amssymb}    

\newcommand{\asec}{$^{\prime\prime}$}
\def\hc3n{HC$_{3}$N}

\begin{document}

   \title{Excitation and spatial study of a prestellar cluster towards G+0.693-0.027 in the Galactic centre} 


   \author{L. Colzi\inst{1}
         \and
         J. Martín-Pintado\inst{1}
          \and
         S. Zeng\inst{2}
          \and
          I. Jiménez-Serra\inst{1}
         \and
          V. M. Rivilla\inst{1}
          \and
          M. Sanz-Novo\inst{1}
          \and
          S. Martín\inst{3,4}
          \and
         Q. Zhang\inst{5}
         \and
           X. Lu\inst{6}
         }

   \institute{Centro de Astrobiología (CAB), CSIC-INTA, Ctra. de Ajalvir Km. 4, 28850, Torrejón de Ardoz, Madrid, Spain \\
              \email{lcolzi@cab.inta-csic.es}
              \and
              Star and Planet Formation Laboratory, Cluster for Pioneering Research, RIKEN, 2-1 Hirosawa, Wako, Saitama, 351-0198, Japan
              \and
              European Southern Observatory, Alonso de Córdova 3107, Vitacura 763 0355, Santiago, Chile
              \and
              Joint ALMA Observatory, Alonso de Córdova 3107, Vitacura 763 0355, Santiago, Chile
              \and
    Center for Astrophysics, Harvard \& Smithsonian, 60 Garden Street, Cambridge, MA 02138, USA
     \and
    Shanghai Astronomical Observatory, Chinese Academy of Sciences, 80 Nandan Road, Shanghai 200030, People's Republic of China
             }

   \date{Received 04 July 2024; Accepted 20 August 2024}

 
  \abstract
   {Star formation in the central molecular zone (CMZ) is suppressed with respect to that of the Galactic disk, and this is likely related to its high turbulent environment. This turbulence impedes the potential detection of prestellar cores. } 
   {We present the temperature, density, and spatial structure of the CMZ molecular cloud G+0.693-0.027, which has been proposed to host a prestellar cluster in the Sgr B2 region.}
   {We analysed multiple HC$_{3}$N rotational transitions that were observed with the IRAM 30m, APEX, Yebes 40m, and GBT radio telescopes, together with SMA+APEX spatially resolved maps.}
   {The spectral shape of HC$_{3}$N lines shows three distinct velocity components: a broad component with a line width of 23 km s$^{-1}$ (C1), and two narrow components with line widths of 7.2 and 8.8 km s$^{-1}$ (C2 and C3). This suggests that a fraction of the molecular gas in this cloud is undergoing turbulence dissipation. From an initial local thermodynamic equilibrium analysis, we found column densities of $N$= (6.54$\pm$0.07)$\times$10$^{14}$ cm$^{-2}$, (9$\pm$3)$\times$10$^{14}$ cm$^{-2}$, and (3.6$\pm$0.7)$\times$10$^{13}$ cm$^{-2}$ for C1, C2, and C3, respectively. These values were used as input for a subsequent non-local thermodynamic equilibrium analysis, in which we found H$_{2}$ densities of 2$\times$10$^{4}$ cm$^{-3}$, 5$\times$10$^{4}$ cm$^{-3}$, and 4$\times$10$^{5}$ cm$^{-3}$ and kinetic temperatures of 140 K, 30 K, and 80 K for C1, C2, and C3, respectively. The spatially resolved maps confirm that the colder and high-density condensations C2 and C3, which  peak in the 70-85 km s$^{-1}$ velocity range, have deconvolved sizes of 9\asec\;(0.36 pc) and 7.6\asec (0.3 pc), respectively, and are embedded in a more diffuse and warmer gas (C1).}
   {The larger-scale structure of the Sgr B2 region, consistently with previous works, shows a hole at 40--50 km s$^{-1}$ that is likely due to a small cloud that shocked the Sgr B2 region and is spatially related with a massive cloud at 60--80 km s$^{-1}$. We propose that the impacting small cloud sequentially triggered the formation of Sgr B2(M), (N), and (S) and the condensations in G+0.693-0.027 during its passage. The two condensations are in a post-shocked environment that has undergone internal fragmentation. Based on the analysis of their masses and the virial parameters, C2 might expand, while C3 might further fragment or collapse.}
   {}


   \maketitle
%

\section{Introduction}

  It is well known that up to 10\% of the new born stars in the Galaxy in the past 100 Myr formed in its central molecular zone (CMZ), that is, in the inner 300 pc. The recent star formation rate (SFR) has been found to be $\sim$0.07-0.09 M$_\odot$/yr, as inferred from the counting of young stellar objects and from the luminosity functions (e.g. \citealt{barnes2017,nandakumar2018,henshaw2023}). 
  However, even though the CMZ contains 80\% of the dense molecular gas in the Galaxy ($M_{\rm CMZ}$=2--6$\times$10$^{7}$ M$_{\odot}$; \citealt{morris1996}), its recent SFR is lower by one order of magnitude than that in the disk (1.5--2 M$_\odot$/yr; e.g. \citealt{licquia2015}). This is not expected from star formation scaling relations, which describe the relation between physical properties of the gas, such as the gas surface density ($\Sigma_{\rm gas}$), and their SFR. \citet{longmore2013} found that most of the gas in the CMZ is above the $\Sigma_{\rm gas}$ threshold of $\sim$ 116 M$_\odot$ pc$^{-2}$ for star formation (\citealt{lada2010,lada2012}), and should follow a linear relation with the SFR. However, it underproduces stars by an order of magnitude with respect to these predictions. 
  
  Only three prominent Galactic centre (GC) massive young stellar clusters exist: the Arches, the Quintuplet, and the Young Nuclear cluster (e.g. \citealt{lu2018}). They account for $<$ 10\% of the expected young stellar mass, as estimated from a massive cluster in the Large Magellanic Cloud (\citealt{schneider2014}). This is known as the missing-cluster problem. 
  
  Concerning the molecular reservoir for star formation,  only $<$ 10\% of the clouds in the CMZ are bound in overdensities that are already forming stars or are on the verge of star formation (\citealt{myers2022,henshaw2023}). The reason for the lack of star-forming clouds in the GC is still unknown. It may be related to the extreme environmental conditions that provide support against gravitational collapse. The high turbulent energy (the typical internal cloud velocity dispersion is $\sim$15-50 km s$^{-1}$; \citealt{morris1996}) increases the critical density threshold for star formation with respect to that in the Galactic disk (e.g. \citealt{kauffmann2013,ginsburg2018,lu2020,tanaka2020}). Moreover, intense ultraviolet (UV) radiation fields produced by stellar clusters, X-rays from the central black hole, enhanced cosmic-ray ionisation rates, and stronger magnetic fields make the physical conditions of star formation in the CMZ different from those in the Galactic disk.

  These conditions do not stop recent star formation entirely, as revealed by the presence of the Arches and Quintuplet young stellar clusters, the Sgr B2 star-forming region associated with several protostars embedded in hot molecular cores (e.g. \citealt{belloche2013,belloche2016,muller2016,bonfand2019,busch2022,busch2024}), the Sgr C star-forming region (e.g. \citealt{forstercaswell2000,kendrew2013,lu2019a,lu2022}), the possible presence of a protostar in Cloud e within the dust-ridge (e.g. \citealt{barnes2018,walker2018}), protostellar outflows in the 20 km s$^{-1}$ cloud (e.g. \citealt{lu2021}), and the G0.253+0.016 molecular cloud, also known as "the Brick", recently proposed as a progenitor of a protocluster (\citealt{walker2021}). Moreover, \citet{peissker2023} found an embedded young stellar cluster near Sgr A$^{*}$, which is the outcome of a recent star formation process. However, the detection of dense (H$_{2}$ density of $n>$10$^{4}$ cm$^{-3}$) and, at the same time, cold\footnote{The typical kinetic gas temperature, $T_{\rm kin}$, of the CMZ is 50-100 K (e.g. \citealt{krieger2017}). The higher the temperature, the higher the Jeans mass of the cloud, and this makes it more stable against gravitational collapse.} gas ($T_{\rm kin}\sim$10-30 K), typical of clouds at early evolutionary stages of star formation, is still elusive in the CMZ. \citet{miettinen2024} recently found that the infrared dark cloud G1.75-0.08 in the CMZ is a candidate high-mass starless cluster in which the clumps are gravitationally unbound.
  It is still unclear whether the non-detection of prestellar sources is directly related to an observational bias due to the presence of warm gas layers with high column densities and high-velocity dispersion, which makes it difficult to distinguish the signatures of the denser and less turbulent prestellar gas. 

  A molecular cloud that stands out in CMZ studies is G+0.693–0.027 (hereafter G+0.693). It lies in the Sgr B2 region and was found to be very prolific from an astrochemical point of view: Many molecules, including complex organic ones, have been detected, some of them for the first time (e.g. \citealt{zeng2019,zeng2021,zeng2023,rivilla2019,rivilla2021a,rivilla2021b,rivilla2023,jimenez-serra2020,jimenez-serra2022,rodriguez-almeida2021a,massalkhi2023,sanz-novo2023,sanandres2024}). G+0.693 is located $\sim$55\asec north-east from Sgr B2(N), a well-known region of ongoing star formation (e.g. \citealt{schmiedeke2016,ginsburg2018}). Conversely, no signpost of star formation, such as masers or continuum sources (e.g. \citealt{ginsburg2018}), was found towards G+0.693. \citet{zeng2020} suggested that this cloud is affected by a cloud-cloud collision that produces the shocks that cause its rich chemistry (e.g. \citealt{requena-torres2006,martin2008,armijos-abendano2020}). \citet{enokiya2022} also suggested that a cloud-cloud collision that occurred $\sim$3 Myr ago created both Sgr B1 and Sgr B2. Sgr B2 harbours three main regions with massive star formation, which are from the south to the north Sgr B2(S), Sgr B2(M), and Sgr B2(N). Sgr B2(N) is thought to be younger than Sgr B2(M), and Sgr B2(S) is younger than Sgr B2(N) (e.g. \citealt{devicente2000,schmiedeke2016,ginsburg2018,jeff2024}). This suggests that the star formation activity occurs sequentially from Sgr B2(M) towards the north and south of this region. In this scenario, the location of G+0.693 makes it a good candidate for future star formation within the Sgr B2 complex. 
  
  In support of this hypothesis, \citet{colzi2022a} recently found that high-$J$ transitions of molecules such as HCO$^{+}$, N$_{2}$H$^{+}$, HCN, HNC, and their less abundant H, C, and N isotopologues towards G+0.693 trace a less turbulent gas (line widths of $\sim$9 km s$^{-1}$) than the typical gas of this source and of the CMZ (line widths of $\sim$20 km s$^{-1}$; e.g. \citealt{requena-torres2006,requena-torres2008,zeng2018,zeng2020}). Moreover, the high D/H ratios of these molecules, together with a radiative transfer analysis, indicated that this gas component has $T_{\rm kin}\leq$30 K, that is, lower than the typical value of 70-140 K (\citealt{krieger2017,zeng2018})\footnote{The dust temperature in the CMZ is $\sim$20 K (\citealt{rodriguez-fernandez2004}), much lower than the typical gas $T_{\rm kin}$.}, and an H$_{2}$ density of $n\geq$5$\times$10$^{4}$ cm$^{-3}$, suggesting the presence of a cold and dense condensation. However, its presence has been inferred indirectly, and we therefore need to clearly establish whether the new condensation is indeed a protocluster on the verge of forming stars. 
  
  In this work, we present a multiple transition analysis of HC$_{3}$N towards G+0.693 to derive gas densities and temperatures, coupled with spatially resolved HC$_{3}$N images of the source. The advantages of studying gas properties using HC$_{3}$N have been highlighted already by several authors (e.g. \citealt{morris1976,morris1977, mills2018}). This molecule presents multiple rotational transitions that lie close to each other (each 9 GHz apart), which allows us to study gas that spans a wide range of critical densities. Moreover, the HC$_{3}$N collisional coefficients are well determined up to 300 K (\citealt{faure2016}), which makes it an ideal tracer for studying the density structure in the CMZ. The paper is organised as follows: In Sect.~\ref{observations} we present the observed data we used in this study. The main results of the analysis are shown in Sect.~\ref{results}  and are discussed in Sect.~\ref{discussion}. Finally, the conclusions are summarised in Sect~\ref{conclusions}.


\section{Observations}
\label{observations}

\begin{table*}
\begin{center}
\caption{\label{table-HC3Ntransitions}HC$_{3}$N transitions in the G+0.693 spectral survey (Sect.~\ref{observations}).}
\begin{tabular}{l l c  l c c c }
\hline
Transition      & Frequency     & log$I$        & $E_{\rm up}$          & Telescope & HPBW & rms\\
($J_{\rm up}$--$J_{\rm low}$) & (GHz)   & (nm$^{2}$ MHz) &  (K)  & & (\asec) & (mK)  \\
\hline
\multicolumn{6}{c}{Low-$J$}\\
\hline
5-4     &45.4903        &-3.164 &4.4    & Yebes 40m & 39 & 0.9\\
8-7     &72.7838        &-2.564 &12.2   & IRAM 30m & 34 & 3.5\\
9-8     &81.8815        &-2.416 &15.7   & IRAM 30m & 30 & 3.5 \\
10-9    &90.9790        &-2.285 &19.6   & IRAM 30m & 27& 1.1\\
11-10   &100.0764       &-2.167 &24.0   & IRAM 30m & 24& 3.3\\
12-11   &109.1736       &-2.061 &28.8   & IRAM 30m & 22& 2.0\\
14-13   &127.3677       &-1.877 &39.7   & IRAM 30m & 19& 9.3\\
15-14   &136.4644       &-1.796 &45.8   & IRAM 30m & 18 & 3.6\\
16-15   &145.5610       &-1.722 &52.4   & IRAM 30m & 17& 4.8\\
17-16   &154.6573       &-1.653 &59.4   & IRAM 30m & 16& 2.3\\
18-17   &163.7533       &-1.590 &66.8   & IRAM 30m & 15& 1.4\\
19-18   &172.8493       &-1.531 &74.7   & IRAM 30m & 14& 1.6\\
\hline
\multicolumn{6}{c}{High-$J$}\\
\hline
22-21   &200.1354       &-1.379 &100.8  & IRAM 30m & 12 & 7.3\\
23-22   &209.2302       &-1.335 &110.5  & IRAM 30m & 12& 7.3\\
\multirow{2}{*}{24-23}\tablefootmark{a} &\multirow{2}{*}{218.3247}      &\multirow{2}{*}{-1.295}        &\multirow{2}{*}{120.5} & IRAM 30m & 11& 7.3\\
&&& & APEX\tablefootmark{b} & 27&1.0 \\
25-24   &227.4189       &-1.257 &131.0  & IRAM 30m & 10& 7.5\\
\multirow{2}{*}{26-25}  &\multirow{2}{*}{236.5128}      &\multirow{2}{*}{-1.222}        &\multirow{2}{*}{141.9} & IRAM 30m & 10 & 7.5\\
& & & & APEX & 25 & 2.1\\
\hline
\multicolumn{6}{c}{Not used in the analysis}\\
\hline
2-1\tablefootmark{c}    &18.1962        &-4.351 &0.4    & GBT & 42 &2.3\\
4-3\tablefootmark{c}    &36.3923        &-3.452 &2.6    & Yebes 40m & 48 & 0.9\\
28-27\tablefootmark{d}    & 254.6995 & -1.160 & 165.0 & IRAM 30m & 10 & 10.6\\
29-28\tablefootmark{e}  &263.7923       &-1.132 &177.2  & APEX & 22 & 1.3\\
30-29\tablefootmark{f}    & 272.8847 & -1.106 & 189.8 & IRAM 30m & 9 & 10.6\\
 \hline
  \normalsize
   \end{tabular}
   \end{center}
\tablefoot{Columns 1 and 2 list the rotational transitions and the related frequencies, respectively. log$I$ is the base 10 logarithm of the integrated intensity at 300 K and $E_{\rm up}$ is the energy of the upper level (Col. 3 and 4). Columns 5, 6, and 7 indicate the telescope used, the relative half-power beam width at the frequency of the transition, and the rms achieved, respectively. \tablefoottext{a}{This transition has also been mapped with SMA+APEX data, as explained in Sect.~\ref{obs-maps};} \tablefoottext{b}{From single-pointing observations described in Sect.~\ref{obs-single-dish}.} \tablefoottext{c}{It includes more extended gas (see Appendix \ref{app-other-transitions}).} \tablefoottext{d}{Contaminated by CH$_{2}$NH.} \tablefoottext{e}{Contaminated by HNCO.} \tablefoottext{f}{Transition intensity within the spectrum noise level.}
}
 \end{table*}

\subsection{Single-dish data}
\label{obs-single-dish}

The data presented in this work are based on a ultrahigh-sensitivity spectral survey obtained towards the G+0.693 molecular cloud carried out with the IRAM\footnote{Institut de radioastronomie millimétrique.} 30m (Granada, Spain), Yebes 40m (Guadalajara, Spain), APEX\footnote{Atacama Pathfinder Experiment telescope.} (Chajnantor, Chile), and GBT\footnote{Green Bank Telescope.} (West Virginia, USA) radio telescopes, and were already used in previous works (e.g. \citealt{rivilla2021a,colzi2022a,jimenez-serra2022,rivilla2022b,massalkhi2023,sanz-novo2023}). The observations, using position-switching mode, were centred at $\alpha_{\rm J2000}$ = 17$^{\rm h}$47$^{\rm m}$22$^{\rm s}$ and $\delta_{\rm J2000}$ = -28$^\circ$21$^{\prime}$27\asec, with an off position shifted by (-885\asec,+290\asec). The line intensity of the spectra was measured in units of antenna temperature corrected for atmospheric attenuation, radiative loss, and rearward scattering and spillover ($T_{\rm A}^{*}$), which is a good assumption based on the extended molecular emission towards G+0.693, beyond the primary beams of the telescopes (e.g. \citealt{brunken2010,jones2012,li2020,zheng2024}). 

The IRAM 30m and Yebes 40m observations were recently obtained by ultradeep spectral surveys that reached sub-mK noise levels. In particular, the IRAM 30m observations are a combination of the projects 172--18 (PI Martín-Pintado; observations made in the period 2019 April 10--16), 018--19 and 133--19 (PI Rivilla; observations made in the periods 2019 August 13--19 and 2019 December 11-15, respectively), and of the new project 123--22 (PI: Jiménez-Serra; observations made in the period 2023 February 1--18). We covered the spectral ranges 71.76--116.72 GHz, 124.77--175.5 GHz, 199.8--238.29 GHz, 252.52--260.30 GHz, and 268.2--275.98 GHz. The final spectra were smoothed to 615 kHz, which translates into a velocity resolution of 1.0--2.2 km s$^{-1}$ in the spectral range covered. This is enough to properly characterise the typical line widths of 15--20 km s$^{-1}$ observed in the region (e.g. \citealt{rivilla2020,rivilla2021a,colzi2022a}). The final noise achieved is 1.7--2.8 mK (at 71--90 GHz), 1.5--9.8 mK (at 90--115 GHz), 3.1--6.8 mK (at 124--175 GHz), 4.7--9.7 mK (at 200--238 GHz), and 10.8--18 mK (at 250--275 GHz). The Yebes 40m observations were obtained during multiple sessions between 2021 March and 2022 March\footnote{Yebes 40m observations were performed in 2021 March 25--21, April 6, 7, 13, 17--19, May 6, 7, 12--16, 18, 20, 21, 26, 30, 31, June 1, December 4--6, 8, 9, 11--13, 15, and 2022 January 26--30, February 3, 18, 19, 20, 22, 26, 27, and March 5, 21, and 22. The final total telescope time scheduled on source was 110 hr.} (project 21A014; PI: Rivilla). We observed a total frequency coverage of 31.07--50.42 GHz. The final spectra were smoothed to 256 kHz, which translates into velocity resolutions of 1.5--2.5 km s$^{-1}$. The achieved noise at this spectral resolution is 0.25--0.9 mK, depending on the spectral range. For more information on these observations, we refer to \citet{rivilla2023} and \citet{sanz-novo2023}.

For the APEX observations (projects O-0108.F-9308A-2021 ad E-0108.C-0306A-2021; PI Rivilla; observations made in the periods 2021 July 11 to September 26 and 2021 October 31 and November 1, respectively) the NFLASH230 receiver connected to two FFTS backends was used, which provide a simultaneous coverage of two side bands of 7.9 GHz each separated by 8 GHz. The spectral resolution was 250 kHz. The covered spectral ranges are 217.93-225.93, 234.18-242.18, 243.94-252.13 and 260.17-268.38 GHz. The final spectra were smoothed to 1 MHz, which translates into a velocity resolution of 1.1-1.4 km s$^{-1}$ in the frequency ranges observed. The achieved noise at this spectral resolution is 1.0--2.2 mK. For further details, we refer to \cite{rivilla2022b}.

The GBT observations were performed in 2009 July-October (PI Requena-Torres). The Ku-band receiver was connected to the spectrometer, providing four 200 MHz spectral
windows with a spectral resolution of 195 kHz, corresponding to a velocity resolution of 2.2–8.6 km s$^{-1}$, and covering a frequency range between 12 GHz and 26 GHz. 

Table \ref{table-HC3Ntransitions} lists the observed \hc3n transitions together with the telescope used, the corresponding half-power beam width (HPBW), and the rms achieved for each transition.

\subsection{High angular resolution data}
\label{obs-maps}

We also used interferometric observations carried out with the Submillimeter array (SMA) at 230 GHz in 2014 as part of the CMZoom survey (\citealt{battersby2020}; see \citealt{zeng2020} for more information). In particular, we used the data centred at the \hc3n(24--23) transition. The spectral resolution was 0.812 MHz, that is, a velocity resolution of $\sim$1.1 km s$^{-1}$, similar to that of the single-dish observations (see Sect.~\ref{obs-single-dish}). The resulting synthesised beam of the maps was $\sim$4\asec.4$\times$4\asec.0 (equivalent to 0.18 pc $\times$ 0.16 pc at the distance of the source, 8.34$\pm$0.16 kpc \citealt{reid2014}) with a position angle of 33$\degree$.7. To recover the missing flux due to the limited uv-coverage of the interferometer, these data were also combined with APEX single-dish data (projects M-091.F-0019-2013 and E-093.C-0144A-2014; PI Ginsburg). We refer to \citet{zeng2020} for detailed information on these observations.

\section{Analysis and results}
\label{results}

\begin{figure*}[h!]
\centering
\includegraphics[width=34pc]{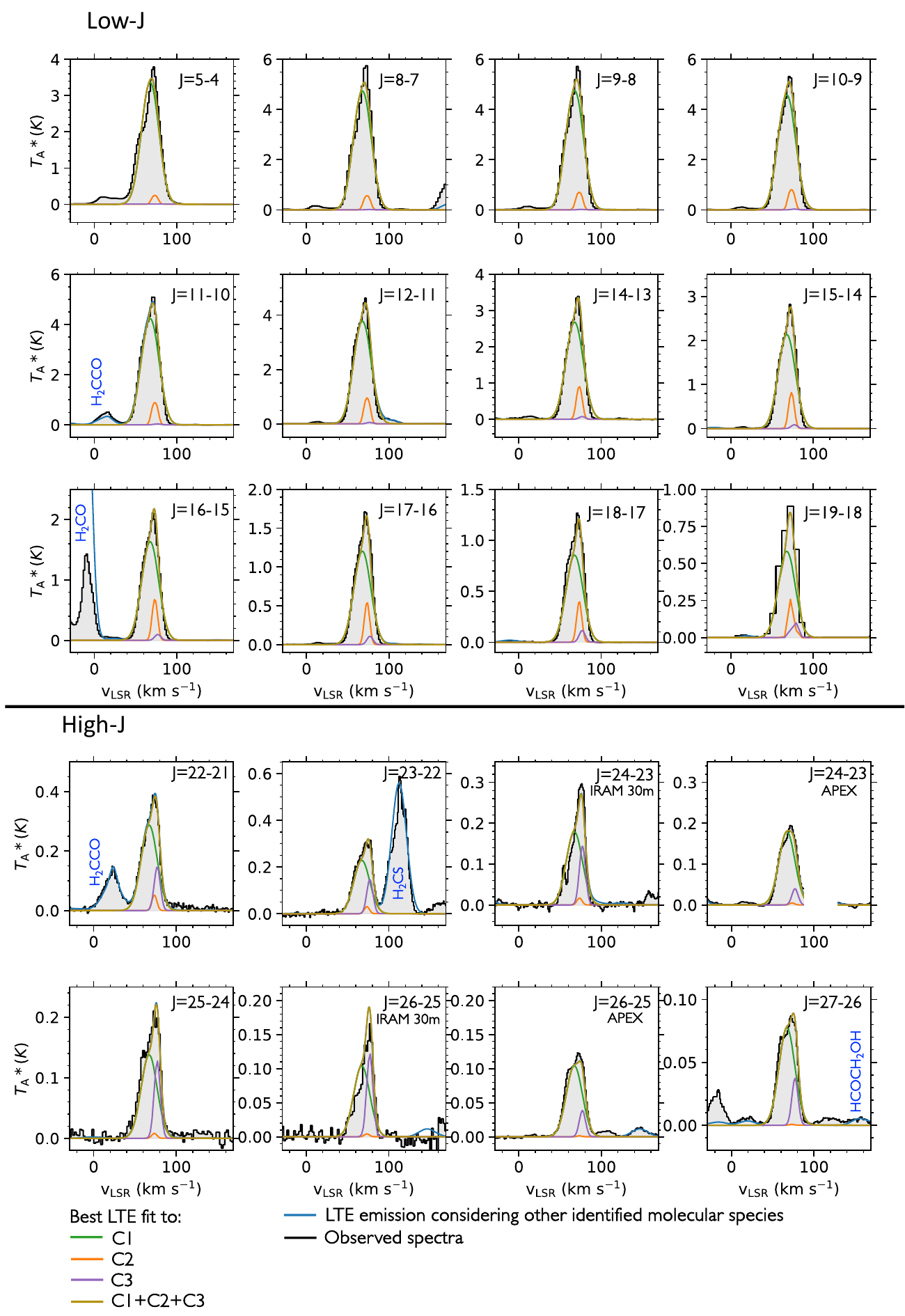}
\caption{HC$_{3}$N observed transitions used for the LTE analysis (black histrograms). The transitions are divided into two groups, the low-$J$ group from $J$=5--4 up $J$=19--18, and the high-$J$ group from $J$=22-21 up to $J$=27-26, in the top and bottom row, respectively. For each panel, the corresponding transition is indicated in the upper right corner, and the telescope information is given when the same transition was observed with more than one instrument. The solid green, orange, and purple lines are the best LTE fit to components C1, C2, and C3, respectively. The dark gold line is the sum of the three components. The blue line indicates the total modelled line emission that also includes the contribution of all molecular species previously identified in the survey (e.g. \citealt{zeng2018,rodriguez-almeida2021a,rodriguez-almeida2021b,rivilla2021a,sanz-novo2023}). More information about the observed transitions is given in Table \ref{table-HC3Ntransitions}.}
\label{fig-hc3n-fit}
\end{figure*}

In Figs.~\ref{fig-hc3n-fit} and \ref{fig-hc3n-nofit} we depict the spectra of the transitions of HC$_{3}$N detected towards G+0.693, whose spectroscopic information is listed in Table \ref{table-HC3Ntransitions} and was taken from the Cologne Database for Molecular
Spectroscopy\footnote{http://cdms.astro.uni-koeln.de/classic/} (CDMS; \citealt{muller2001,muller2005}; \citealt{endres2016}). The entry of the catalogue is based on the laboratory works of \citet{dezafra1971}, \citet{creswell1977}, \citet{mallinson1978}, \citet{chen1991}, \citet{yamada1995}, and \citet{thorwirth2000}. The dipole moment was determined by \citet{deleon1985}. The 28--27, 29--28, and 30--29 rotational transitions are included within the observed spectral setup, but are contaminated by other molecular species or their emission is within the noise, and thus, they were not included in the analysis. 

In this section, we present the multi-transition analysis of HC$_{3}$N both under local thermodynamic equilibrium (LTE) and non-LTE approximations, complemented by a study at high angular resolution (Sects.~\ref{res-LTE}, \ref{res-maps}, and \ref{res-non-LTE}).

\subsection{Local thermodynamic equilibrium analysis: Determining excitation temperatures and column densities}
\label{res-LTE}

\begin{figure}
\centering
\includegraphics[width=20pc]{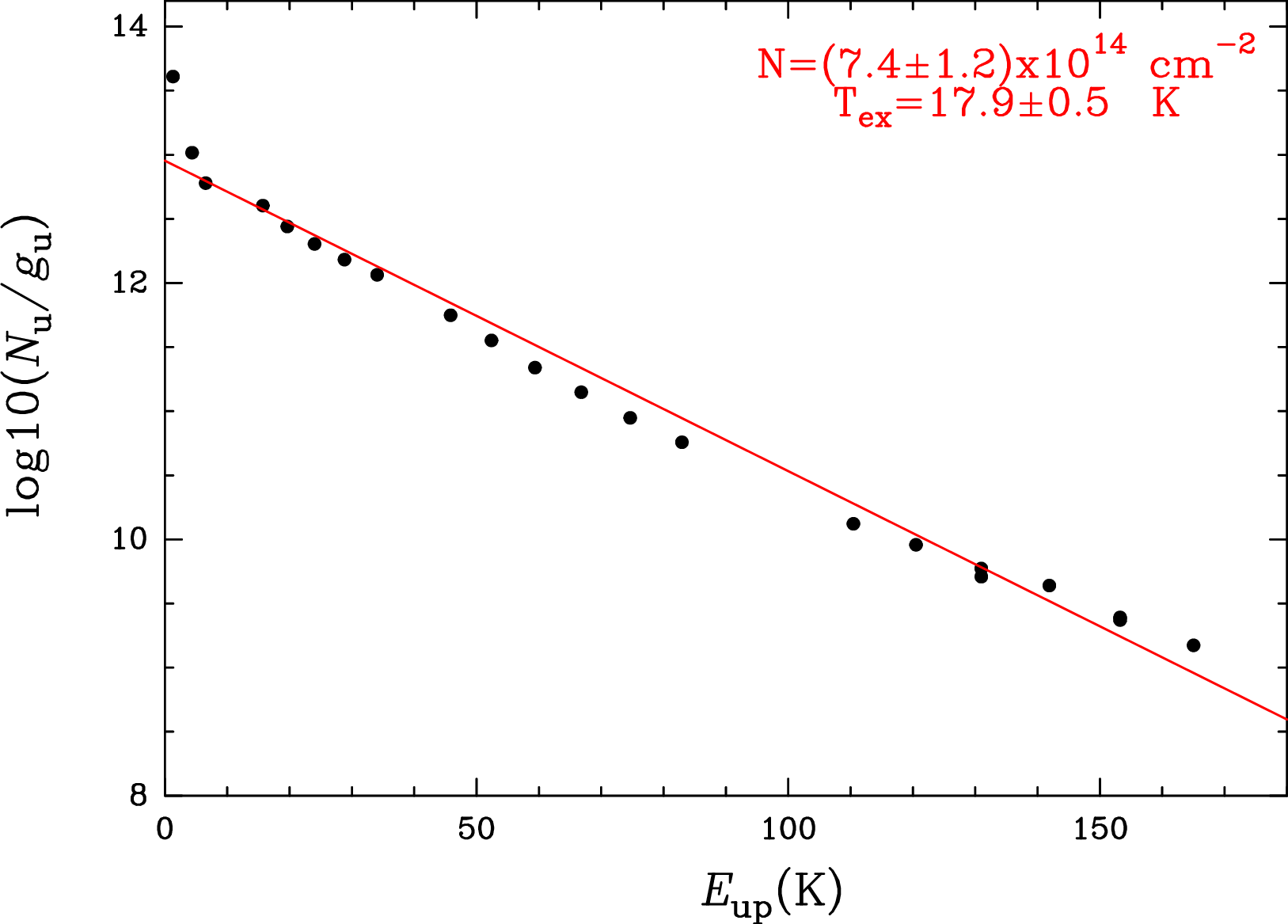}
\caption{Rotational diagram for all the transitions shown in Fig.~\ref{fig-hc3n-fit} and listed in Table \ref{table-HC3Ntransitions} (except for 28--27, 29--28, and 30--29). The solid red line represents a linear regression fit of the points, and the final $N$ and $T_{\rm ex}$ are shown in the upper right corner.} 
\label{fig-rotdiag}
\end{figure}

\begin{figure}
\centering
\includegraphics[width=20pc]{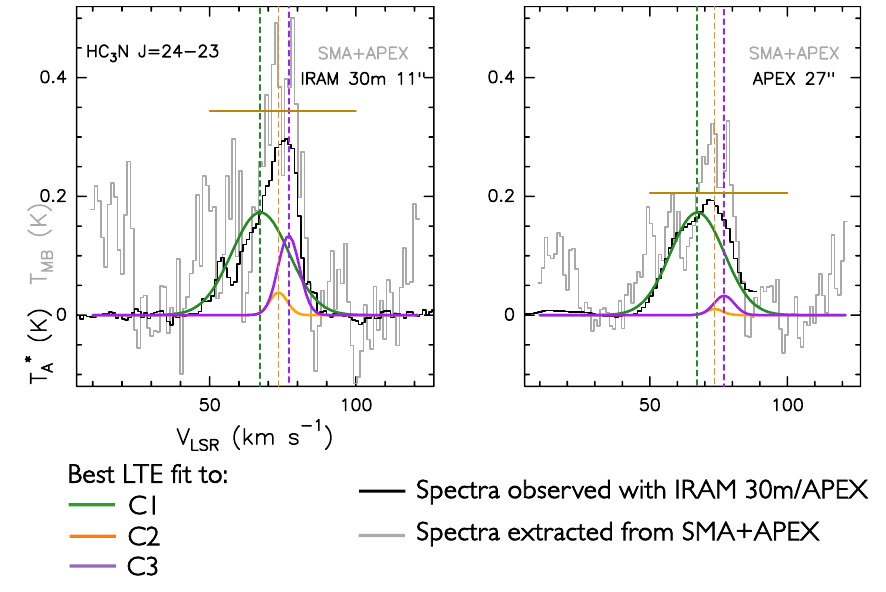}
\caption{Comparison between single-pointing single-dish spectra and spectra extracted from SMA+APEX data cubes. \textit{Left panel}: HC$_{3}$N(24--23) spectrum observed with IRAM 30m (black histogram) together with the LTE predictions of the three components (see Fig.~\ref{fig-hc3n-fit} for more information about the coloured lines). \textit{Right panel}: Same as the left panel, but for the APEX spectrum. In both panels, the spectrum extracted from the HC$_{3}$N(24--23) SMA+APEX cube (grey histograms) is shown for comparison in a region equal to the IRAM 30m beam of 11\asec\,and to the APEX beam of 27\asec (left and right panels, respectively), and centred on G+0.693 (in $T_{\rm MB}$; see Sect.~\ref{res-maps} for more information about the maps). In both panels, the solid dark gold horizontal line indicates the peak of the simulated LTE spectrum, obtained by converting the narrow line peaks (components C2 and C3) from $T_{\rm A}^{*}$ to $T_{\rm MB}$. The dashed vertical lines correspond to the $\varv_{\rm LSR}$ of the three components (see Table \ref{table-result} for high-$J$ transitions).}
\label{fig-spectrasinglevsmap}
\end{figure}

\begin{figure*}
\centering
\includegraphics[width=20pc]{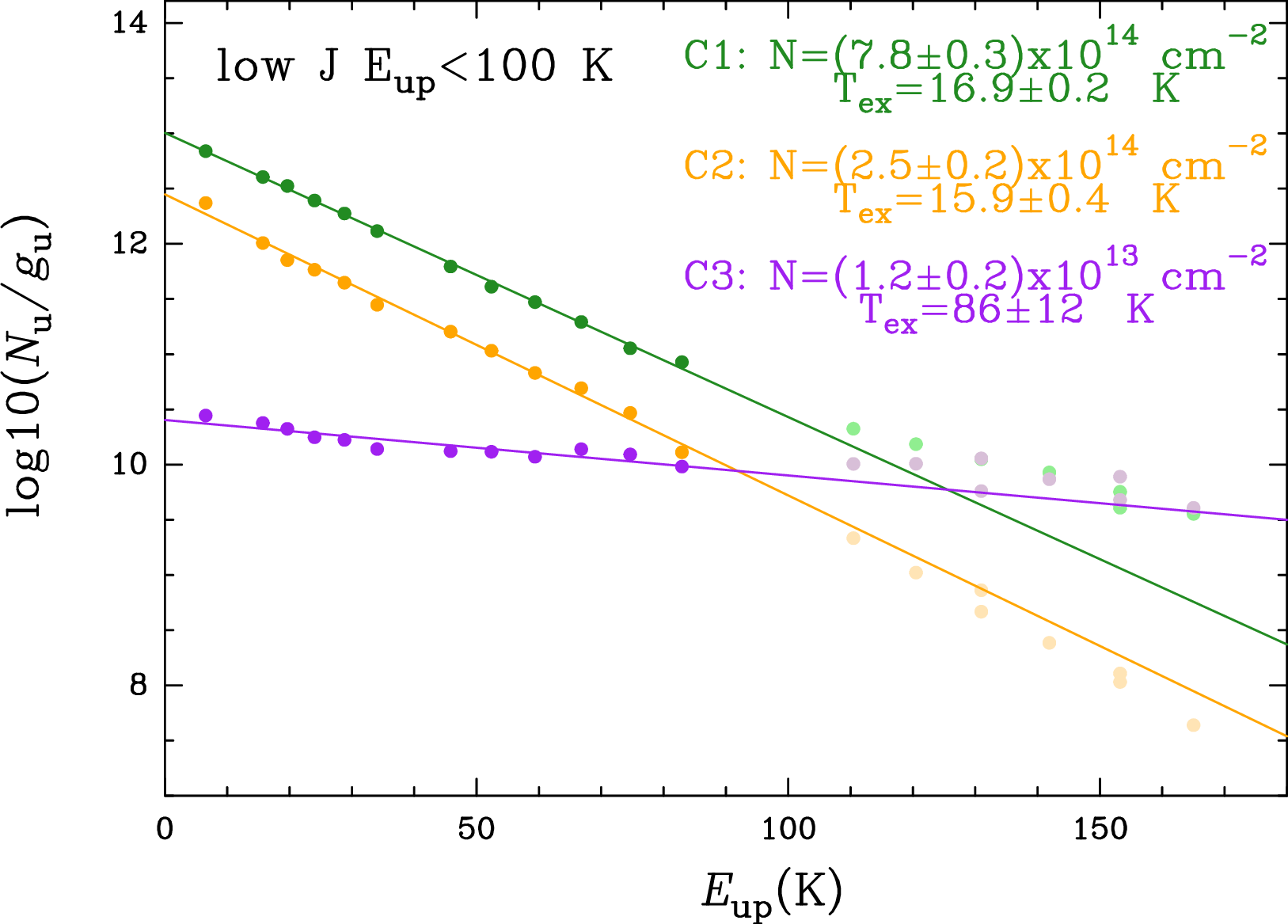}
\hspace{0.5cm}
\includegraphics[width=20pc]{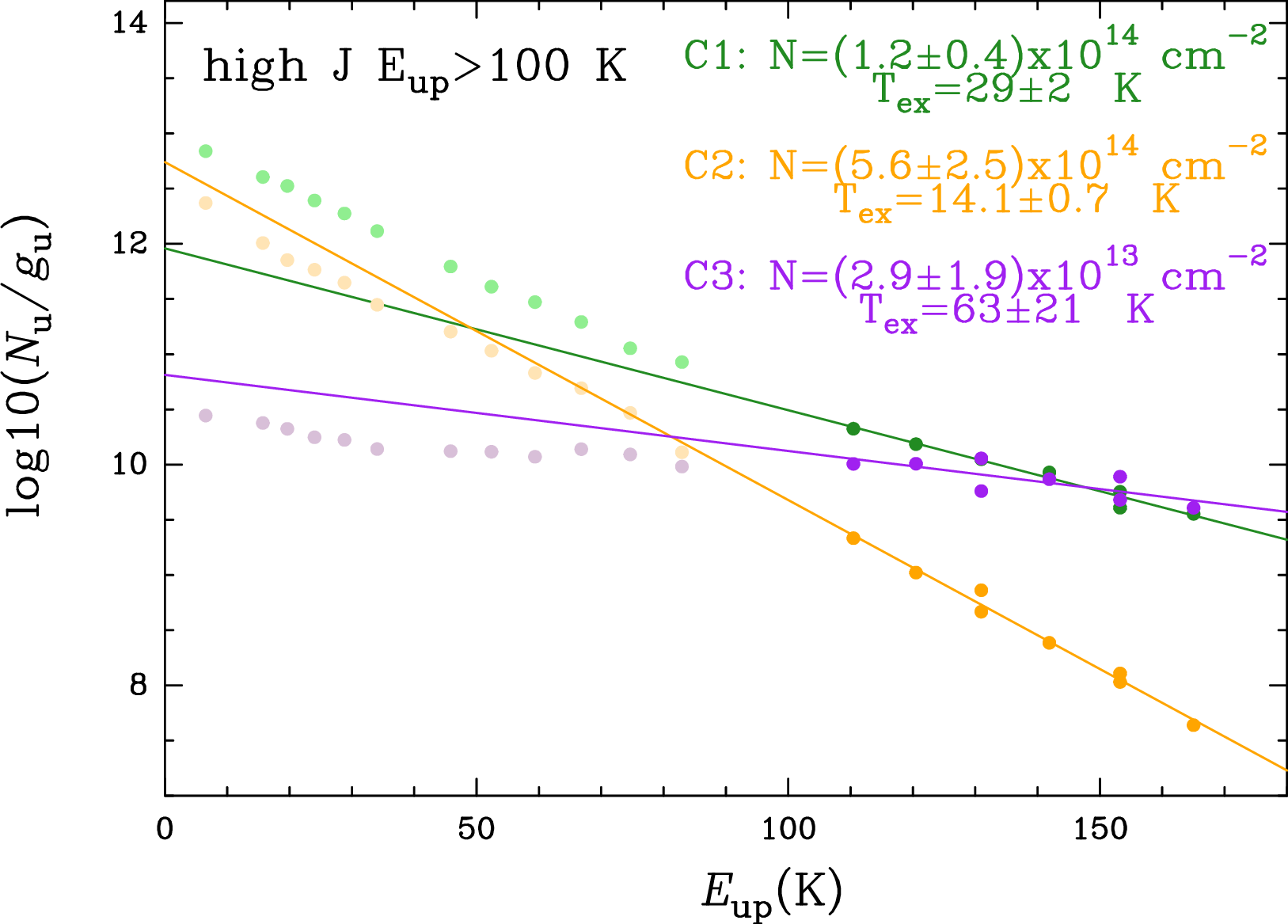}
\caption{Rotational diagram from the low-$J$ (with solid points in the \textit{left panel} with $E_{\rm up}<$ 100 K) and high-$J$ (with solid points in the \textit{right panel} with $E_{\rm up}>$ 100 K) transitions shown in Fig.~\ref{fig-hc3n-fit} and listed in Table \ref{table-HC3Ntransitions}, taking the contributions of the three line components into account. The green, orange, and purple points correspond to components C1, C2, and C3, respectively. The solid coloured lines represent a linear regression fit of the points corresponding to the same component, and the final $N$ and $T_{\rm ex}$ are shown in the upper right corner. For the low-$J$ rotational diagram, the high-$J$ points are shown with lighter colours for completeness, and vice versa.}
\label{fig-rotdiag2}
\end{figure*}

\begin{table*}
\begin{center}
\caption{\label{table-result} Results from the multicomponent HC$_{3}$N LTE fitting procedure and from the rotational diagram (rd).}
\begin{tabular}{lccccccc}
\hline
Component & FWHM        & $\varv_{\rm LSR}$     & $T_{\rm ex}$ & $N$ & $T_{\rm ex}^{\rm (rd)}$ & $N^{\rm (rd)}$        & $\theta_{\rm source}$\tablefootmark{a} \\
& (km s$^{-1}$) & (km s$^{-1}$)   & (K)  &  ($\times$10$^{14}$ cm$^{-2}$) & (K)  & ($\times$10$^{14}$ cm$^{-2}$)& (\asec / pc)\\
\hline
\multicolumn{8}{c}{low-$J$ transitions ($E_{\rm up}<$ 100 K)}\\
\hline
C1 - broad & 23.62$\pm$0.17 & 67.79$\pm$0.09 & 16.53$\pm$0.15 & 6.54$\pm$0.07\tablefootmark{b} & 16.9$\pm$0.2 & 7.8$\pm$0.3& $>$27 / $>$1\tablefootmark{c}\\
C2 - narrow & 7.2$\pm$0.6  & 73.5\tablefootmark{c}  &14.5$\pm$1.7 &  9$\pm$3\tablefootmark{b} &  15.9$\pm$0.4& 2.5$\pm$0.2& 9 / 0.36\tablefootmark{c}\\
C3 - narrow & 8.8\tablefootmark{c} & 77\tablefootmark{c} &  64\tablefootmark{c}  & 0.29$\pm$0.18 &  86$\pm$12&  0.12$\pm$0.02& 7.6 / 0.3\tablefootmark{c}\\
\hline
\multicolumn{8}{c}{high-$J$ transitions ($E_{\rm up}>$ 100 K)}\\
\hline
C1 - broad & 22.8$\pm$0.4 & 67.2$\pm$0.3 & 31.5$\pm$0.8 & 0.93$\pm$0.08 &  29$\pm$2 & 1.2$\pm$0.4&$>$27 / $>$1\tablefootmark{c} \\
C2 - narrow & 7.2\tablefootmark{c} & 73.5\tablefootmark{c} & 14.5\tablefootmark{c} &   7.1$\pm$1.7 &  14.1$\pm$0.7 & 5.6$\pm$2.5& 9 / 0.36\tablefootmark{c}\\
C3 - narrow & 8.8$\pm$0.5& 77\tablefootmark{c} & 64$\pm$10  & 0.36$\pm$0.07\tablefootmark{b} &  63$\pm$21 & 0.29$\pm$0.19& 7.6 / 0.3\tablefootmark{c}\\
\hline
  \normalsize
   \end{tabular}
   \end{center}
\tablefoot{$^{\rm (rd)}$ Results from the rotational diagram. \tablefoottext{a}{The source size corresponds to the diameter;} \tablefoottext{b}{Used as input for the non-LTE analysis explained in Sect.~\ref{res-non-LTE}; \tablefoottext{c}{Parameters without errors are fixed in the fitting procedure, as explained in Sect.~\ref{res-LTE}.}}
   }
 \end{table*}

First, we performed an LTE analysis of the HC$_3$N rotational transitions we observed. They are listed in Table \ref{table-HC3Ntransitions}. They are free of the emission of other molecular species. 
As first step, we used the rotational diagram (see e.g. \citealt{goldsmith1999}) that is implemented in the tool called spectral line identification and modeling (\texttt{SLIM}) within the \texttt{MADCUBA} package\footnote{Madrid Data Cube Analysis on ImageJ is a software program developed at the Center of Astrobiology (CAB) in Madrid; \url{https://cab.inta-csic.es/madcuba/}.} (\citealt{martin2019,rivilla2021a,sanandres2024}), shown in Fig.~\ref{fig-rotdiag}. The results from this analysis are a total column density, $N$, of (7.4$\pm$1.2)$\times$10$^{14}$ cm$^{-3}$, and an excitation temperature, $T_{\rm ex}$, of 17.9$\pm$0.5 K. However, the rotational diagram clearly shows that not all the transitions can be properly fitted with a single straight line. A single LTE simulated spectrum is not able to properly reproduce the observed line profiles. 

The lowest energy transitions, $J$=2--1 and 4--3 (with $E_{\rm up}<$ 5 K), appear to be clear outliers in the rotational diagram (Fig.~\ref{fig-rotdiag}). They are located well above the linear fit.  This is likely because these low-excitation transitions trace a significantly more extended gas. The observed line profiles differ from those of the other transitions (compare Fig.~\ref{fig-hc3n-nofit} with Fig.~\ref{fig-hc3n-fit}), and the line intensities are higher than those predicted by the LTE model, which suggests additional gas components. 
Therefore, we do not use these two low-energy transitions of \hc3n in our analysis below.

To obtain a proper fit, the remaining transitions were divided into two sub-groups: those with 5 $<E_{\rm up}\leq$ 100 K (low-$J$, from 5--4 up to 19--18), and those with $E_{\rm up}>$ 100 K (high-$J$, from 22--21 up to 27--26). As we discuss below (Sect. \ref{res-non-LTE}), this is due to non-LTE effects of the HC$_{3}$N emission towards this source.

To perform an LTE fit to the two sub-groups of transitions, we used the \texttt{SLIM} tool within \texttt{MADCUBA}, which generates a synthetic spectrum assuming LTE conditions and considering the line opacity.
The simulated LTE spectra in \texttt{SLIM} perform the convolution of a Gaussian source distribution with the Gaussian beam for every transition (Table \ref{table-HC3Ntransitions}; for more details, see Sect. 3.2.2 in \citealt{martin2019}).
To derive the physical parameters describing the molecular emission, we used the automatic fitting routine \texttt{AUTOFIT}, which provides the best non-linear least-squares LTE fit to the data using the Levenberg–Marquardt algorithm (\citealt{martin2019}). The free parameters considered in the fit are the column density of the molecule, $N$, the excitation temperature, $T_{\rm ex}$, the velocity, $\varv_{\rm LSR}$, the full width at half maximum (FWHM), and the source size ($\theta_{\rm source}$).
Fig.~\ref{fig-hc3n-fit} shows that most of the observed lines cannot be fitted with a single-Gaussian component. This behaviour is similar to what was found by \citet{colzi2022a} from the D-, $^{13}$C-, $^{15}$N-, and $^{18}$O-isotopologues of HCN, HNC, HCO$^{+}$ and N$_{2}$H$^{+}$, where these lines were fitted with two emission components. 

To properly fit the HC$_{3}$N emission, we found that three distinct velocity components are needed. The first component is at $\sim$67.5 km $^{-1}$, with a FWHM of $\sim$23 km s$^{-1}$ (green line in Fig.~\ref{fig-hc3n-fit}), which belongs to the G+0.693 warm and turbulent gas. This component (hereafter, component C1) usually dominates the emission of most of molecular species (\citealt{zeng2018}; \citealt{rivilla2021a}; \citealt{sanz-novo2023}). 

The second component, at $\sim$73.5 km s$^{-1}$, corresponds to the one found by \citet{colzi2022a}, with a narrower FWHM of $\sim$7 km s$^{-1}$. This component (hereafter, component C2) clearly appears in low-$J$ transitions as a narrow peak at higher velocities (orange line in Fig.~\ref{fig-hc3n-fit}). 

Moreover, a third narrow component (hereafter, component C3) appears in high-$J$ lines at even higher velocities, $\sim$77 km s$^{-1}$ (purple line in Fig.~\ref{fig-hc3n-fit}). 
The relevance of this third component increases with the energy of the transition (see the transitions observed with the IRAM 30m telescope in Fig.~\ref{fig-hc3n-fit}), and it is needed to properly reproduce the line shape of the high-$J$ transitions.

The fitting procedure required a total of 15 free parameters to take the three velocity components into account. Thus, to help the fit converge, we fixed some of them as explained below. The $\varv_{\rm LSR}$ for C2 and C3 was fixed to the values that best reproduced the observed spectra, 73.5 km s$^{-1}$ and 77 km s$^{-1}$, respectively. The FWHM and $T_{\rm ex}$ of C2 for high-$J$ transitions were fixed to 7.1 km s$^{-1}$ and 14.5 K, which were derived from the fit of the low-$J$ transitions, where the component is more dominant. In a similar way, the FWHM and $T_{\rm ex}$ of C3 for low-$J$ transitions were fixed to 8.6 km s$^{-1}$ and 64 K, which were derived from the fit of the high-$J$ transitions.

To understand the source extent, $\theta_{\rm source}$, of the different components, we inspected the spectra of the transitions $J$=24--23 and 26--25, which were observed with both the IRAM 30m and APEX radio telescopes. Consequently, the observations of the two telescopes have different beam sizes (see Table \ref{table-HC3Ntransitions}). Therefore, if the emission of a component was extended, its line intensity was expected to be the same observed with both telescopes; while if the emission was compact, it has different beam dilutions, and hence their line intensities are expected to be different. The inspection of the  spectra of the $J$=24--23 and 26--25 transitions (Figs. \ref{fig-hc3n-fit}, \ref{fig-spectrasinglevsmap}, and \ref{fig-hc3n30mapex}) showed that while the line shape and intensity of the broad component C1 seen with both telescopes is similar (indicating that it is extended), the narrow-line peak at higher velocities (mainly produced by C3 for these two high-$J$ transitions) is significantly more intense in the smaller IRAM 30m beam. This indicates that this component is compact, because it is more diluted in the larger APEX beam.

Therefore, for the line fitting, we considered that component C1 is extended, and hence, no beam dilution was applied. Regarding the narrow component C3, we estimated its source size (the diameter),  $\theta_{\rm source}$, using two complementary methods. 
First, we compared the spectral shape of the transitions $J$=24--23 and 26--25 observed with the IRAM 30m and APEX telescopes. 
For details of the size estimation considering the different beam dilutions, we refer to Appendix \ref{app-sourcesize}. With this method, we found a source size of 7.6\asec\,for component C3. 
Complementarily, we measured deconvolved source sizes of 9\asec\,and 7.6\asec\,for components C2 and C3, respectively, using the high angular resolution SMA observations (for more details, see the spatial analysis in Sect.~\ref{res-maps}). This second method confirmed the source size we obtained for C3 from the spectral shapes described above.

The final column densities were calculated using the $T_{\rm A}^{*}$ scale for the broad and extended component C1, and using the $T_{\rm MB}$ scale for the two narrow and compact components C2 and C3. $T_{\rm MB}$ was evaluated via the relation $T_{\rm A}^{*}$ = $T_{\rm MB}\eta_{\rm MB}$, where $\eta_{\rm MB}$ is the ratio of the main beam efficiency ($B_{\rm eff}$) and the forward efficiency ($F_{\rm eff}$) of the telescope. 
Table \ref{table-result} shows the results from the fitting procedure. The $N$ and $T_{\rm ex}$ derived for the broad component C1 of low-$J$ transitions are similar to those derived from the rotational diagram shown in Fig.~\ref{fig-rotdiag}, with all the rotational transitions and without the narrow components. This highlights the fact that to retrieve the total column density of the extended, warm, and turbulent gas component usually studied towards this source (e.g.~\citealt{zeng2018,rivilla2019,sanandres2023}), an LTE analysis of a single component gives good results.

We also performed a rotational diagram analysis of the separated velocity components and transitions using \texttt{MADCUBA}. The calculated integrated intensity took the overlap of the line components into accout using the peak intensity ratios between them and assuming that the emission has Gaussian line profiles, with the line widths and the radial velocities obtained from the \texttt{SLIM} fit.
The results are shown in Fig.~\ref{fig-rotdiag2} and listed in Table \ref{table-result}. $T_{\rm ex}$ and $N$ are consistent within 3$\sigma$ (in the worst case) with those obtained with the LTE procedure fit.

\subsection{HC$_{3}$N spatial distribution}
\label{res-maps}

\begin{figure}
\centering
\includegraphics[width=20pc]{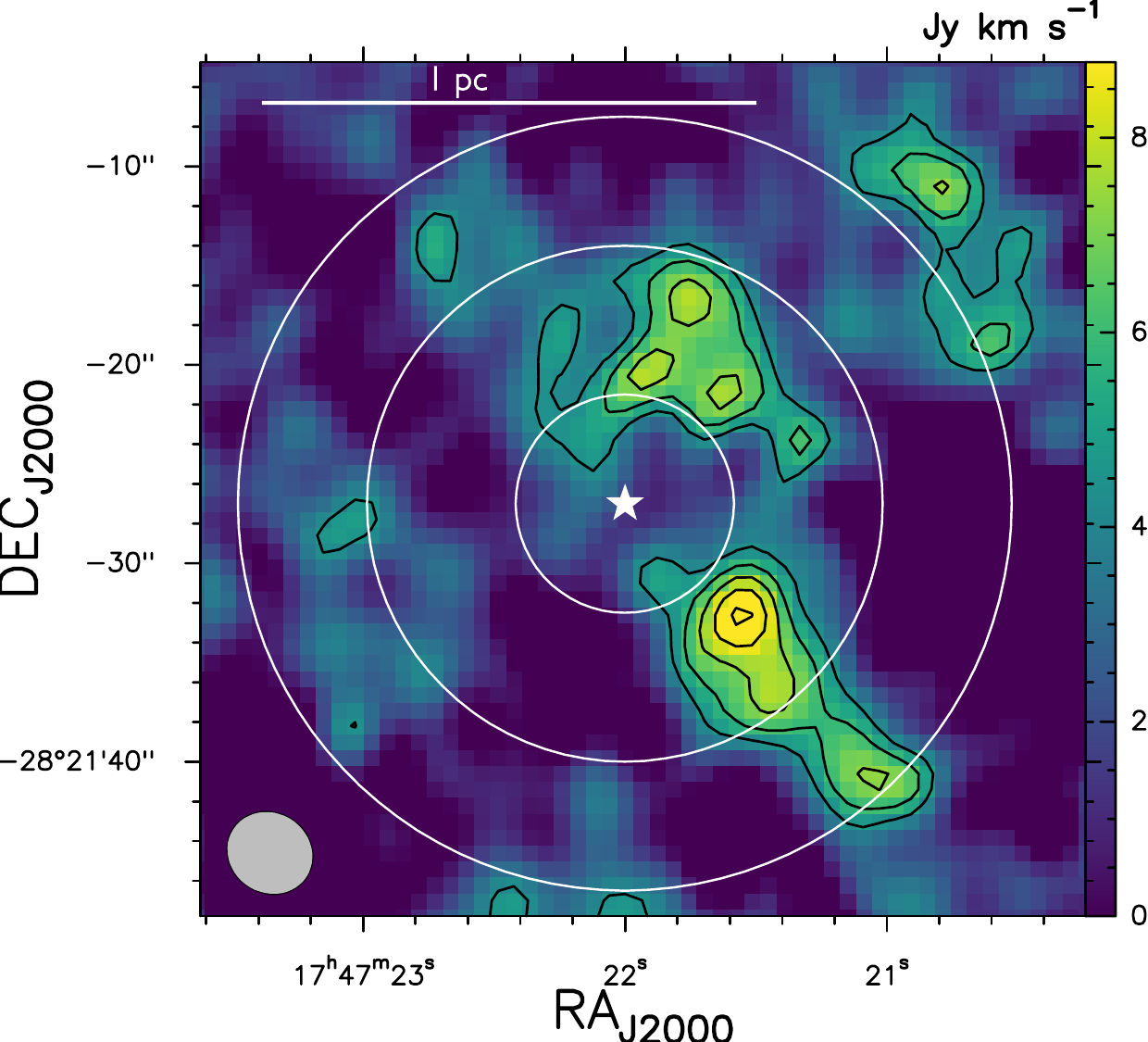}
\caption{HC$_{3}$N(24--23) integrated-intensity map in the velocity range 56-80 km s$^{-1}$. The contour levels are 3, 4, 5, and 6 times the 1$\sigma$ rms of the map, equal to 1.215 Jy km s$^{-1}$. The white star indicates the position of G+0.693 towards which single-dish observations are centred (C1). The white circles show the IRAM 30m and APEX beams at 218.324 GHz (11\asec\,and 27\asec, respectively), and the maximum beam of the single-dish observations studied in this work, which is 39\asec\;from the Yebes beam at 45.490 GHz (HC$_{3}$N(5--4)). The interferometric synthesised beam (4\asec.4$\times$4\asec.0) is the ellipse indicated in the lower left corner. In the top part, the solid white line indicates the spatial scale of 1 pc.}
\label{fig-hc3n_24_23_all}
\end{figure}

\begin{figure*}
\centering
\includegraphics[width=43pc]{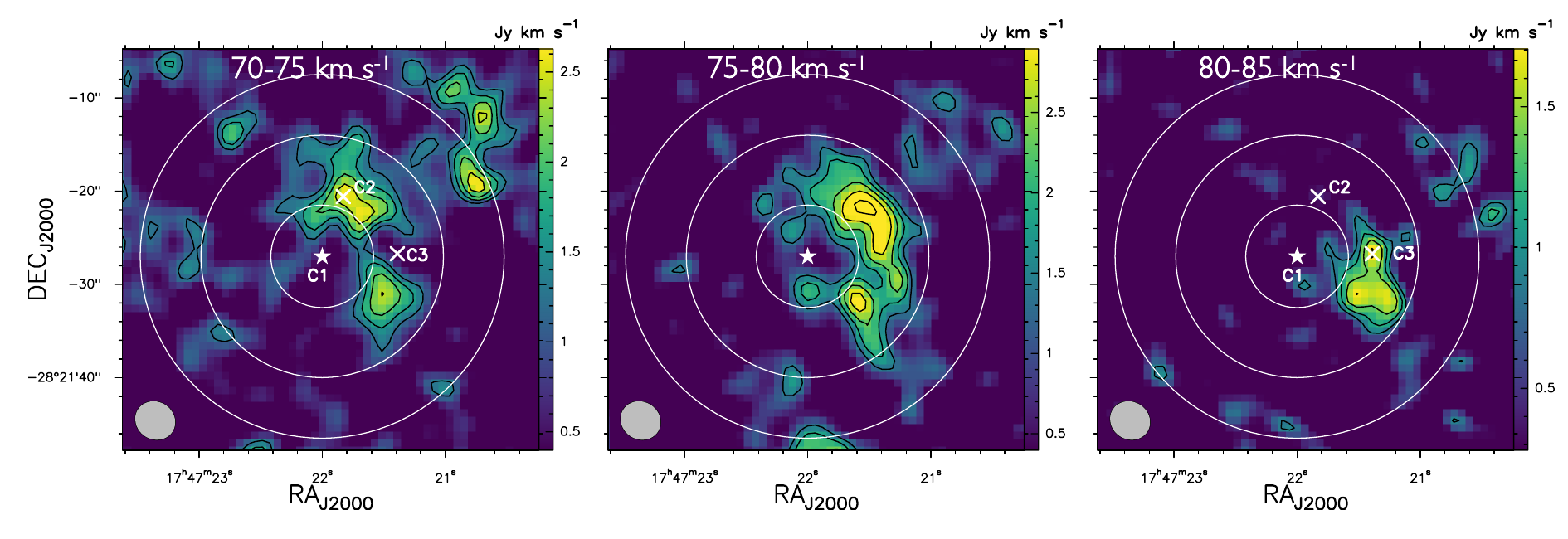}
\caption{Same as Fig.~\ref{fig-hc3n_24_23_all} in the velocity ranges 70-75 km s$^{-1}$ (\textit{left panel}), 75-80 km s$^{-1}$ (\textit{central panel}), and 80-85 km s$^{-1}$ (\textit{right panel}). The 1$\sigma$ rms is 0.397, 0.397, and 0.280 Jy km s$^{-1}$ for the three maps, respectively. Positions C2 and C3 are indicated with white crosses in the left and right panels, whose coordinates are listed in Table \ref{table-coordinates-nonLTE}.}
\label{fig-channelmaps}
\end{figure*}

\begin{figure*}
\centering
\includegraphics[width=30pc]{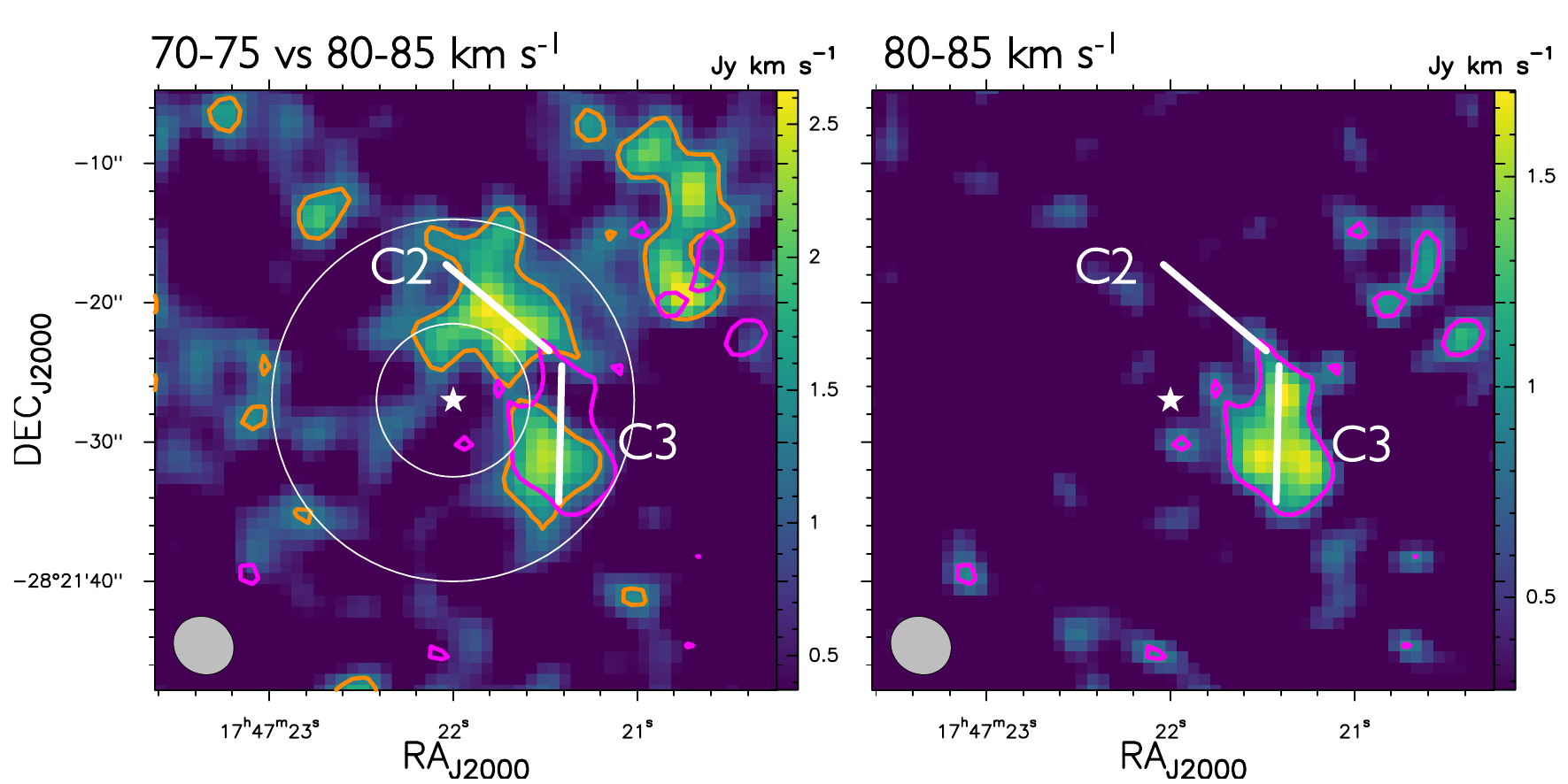}
\includegraphics[width=10pc]{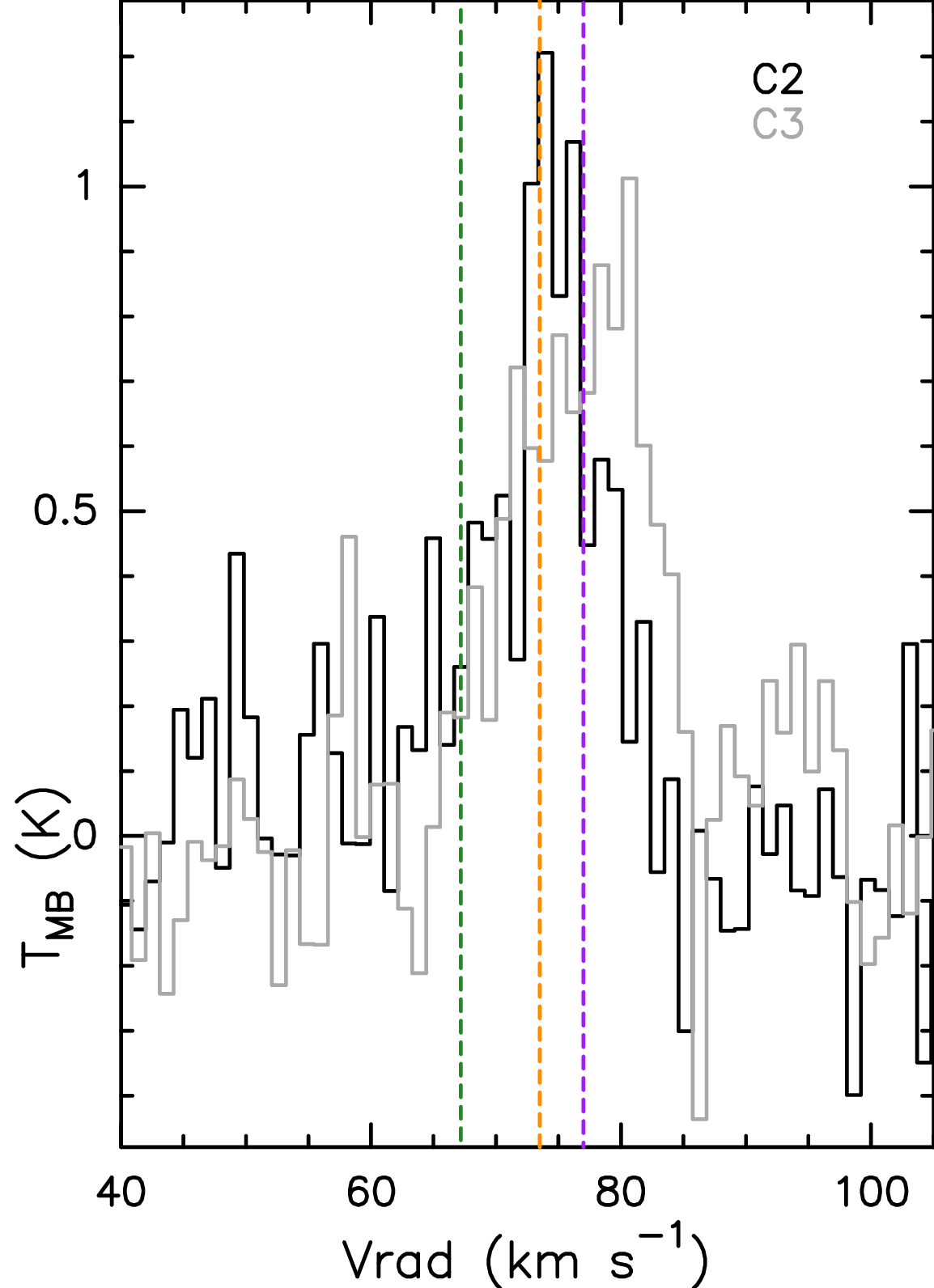}
\caption{Identification of the condensations C2 and C3 from integrated-intensity maps and their extracted spectra. \textit{Left and central panels}: Colours show the HC$_{3}$N(24--23) integrated-intensity maps in the velocity channels 70--75 and 80-85 km s$^{-1}$. The orange contours correspond to half of the peak value of the 70--75 km s$^{-1}$ map (1.7 Jy km s$^{-1}$), and purple contours correspond to half of the peak value the 80--85 km s$^{-1}$ map (2.6 Jy km s$^{-1}$). The solid white lines represent the directions of the pv diagrams shown in Fig.~\ref{fig-pv-comp2-comp3}. \textit{Right panel}: Spectrum extracted from the C2 condensation defined by the orange contours in the north (black histogram), and the spectrum extracted from the C3 condensation defined by the purple contours (grey histogram). The dashed vertical green, orange, and purple lines indicate the $\varv_{\rm LSR}$ of the three components obtained from the LTE analysis (see Sect.~\ref{res-LTE}).}
\label{fig-map-2velcomp}
\end{figure*}

\begin{figure}
\centering
\includegraphics[width=22pc]{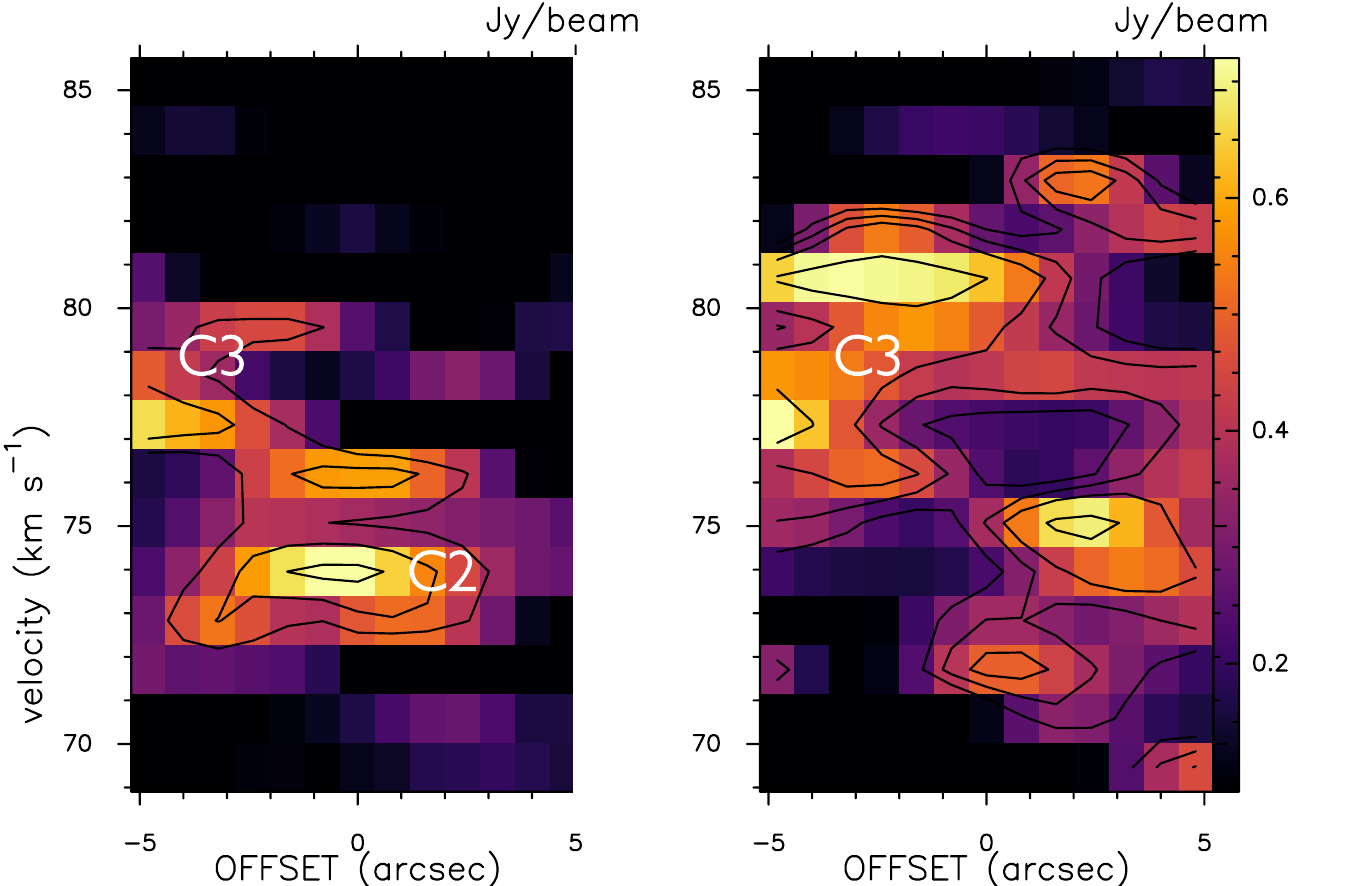}
\caption{Position velocity diagrams along the C2 and C3 condensations (\textit{left} and \textit{right} panels, respectively), following the white lines defined in Fig.~\ref{fig-map-2velcomp}.}
\label{fig-pv-comp2-comp3}
\end{figure}

\begin{table}
\setlength{\tabcolsep}{5pt}
\begin{center}
\caption{\label{table-coordinates-nonLTE} Coordinates and results from the non-LTE analysis of the three velocity components towards G+0.693.}
\begin{tabular}{lcccc}
\hline
Component & RA(J2000) &  Dec(J2000) &$n_{\rm H_{2}}$    & $T_{\rm{kin}}$ \\
& (h m s) & ($\degree\;\prime\;\prime\prime$)& (cm$^{-3}$) & (K) \\
\hline
C1 \tablefootmark{a} & 17:47:22.00 & -28:21:27.00 & 2$\times$10$^{4}$ & 140  \\
C2 & 17:47:21.83   &    -28:21:20.57& 5$\times$10$^{4}$ & 30 \\
C3 & 17:47:21.39 &  -28:21:26.71& 4$\times$10$^{5}$ & 80 \\
\hline
\normalsize
\end{tabular}
\end{center}
\tablefoot{\tablefoottext{a}{Single-dish observations are pointing towards these coordinates.}
}
 \end{table}

\begin{table}
\setlength{\tabcolsep}{4pt}
\begin{center}
\caption{\label{table-grid} Grids of H$_{2}$ density and kinetic temperature models used as input to \texttt{RADEX} for the three components, as explained in Sect.~\ref{res-non-LTE}.}
\begin{tabular}{l ll}
\hline
Component & $n_{\rm H_{2}}$ grid ($\times$10$^{4}$ cm$^{-3}$)   &  $T_{\rm kin}$ grid (K)   \\
\hline
\multirow{2}{*}{C1}   &  \multirow{2}{*}{1, 2, 3, 4, 5, 6, 7}   & 100, 110, 120, 130,    \\
& & 140, 150, 160\\
\hline
\multirow{3}{*}{C2}  &  1, 2, 3, 4, 5, 6, 7, & \multirow{2}{*}{10, 20, 30, 40, 50, 60,}  \\
    &  8, 9, 10, 20, 30, 40,& \multirow{2}{*}{70, 80, 90, 100} \\
    & 50, 60, 70, 80, 90, 100&\\
\hline
\multirow{3}{*}{C3} & 1, 2, 3, 4, 5, 6, 7,   & \multirow{3}{*}{50, 60, 70, 80, 90, 100}  \\
    & 8, 9, 10, 20, 30, 40,   \\
    &50, 60, 70, 80, 90, 100 & \\
\hline
\normalsize
\end{tabular}
\end{center}
 \end{table}

The HC$_{3}$N(24--23) integrated-intensity map in the velocity ranges 56--80 km s$^{-1}$ obtained with the SMA ($+$APEX) data described in Sect.~\ref{obs-maps} is shown in Fig.~\ref{fig-hc3n_24_23_all}. This velocity range corresponds to 68$\pm$12 km s$^{-1}$, where 68 km s$^{-1}$ is the velocity of component C1, and $\sim$12 km s$^{-1}$ is its FWHM/2, which also includes components C2 and C3. Extracting the spectra from the interferometric cube smoothed with \texttt{MADCUBA} to the same beam sizes of the IRAM 30m and APEX, centred at the G+0.693 coordinates, and comparing them with the single-dish observations, we find that the flux is entirely recovered (see Fig.~\ref{fig-spectrasinglevsmap}). The higher angular resolution maps allow us to determine the spatial origin of the narrow components found in the single-dish observations.

To study the spatial origin of the emission of the narrow components C2 and C3 in detail, we produced maps for different velocity ranges. The low-velocity emission (50--70 km s$^{-1}$; see Fig.~\ref{fig-vel-50-60}) mainly comes from the south-west (SW), and can be attributed to the blue-shifted emission of the cloud-cloud collision towards G+0.693 (\citealt{zeng2020}), while the higher-velocity emission ($>$70 km s$^{-1}$) comes from the red-shifted cloud. From now on, we do not discuss the blue-shifted emission because it is severely contaminated in the single-dish observations by the broad turbulent and more extended component C1. The velocities 70--85 km s$^{-1}$ correspond to those obtained from the LTE analysis for the narrower emission of components C2 and C3 (see Fig.~\ref{fig-vel-70-85}). 
To retrieve their spatial origin, we produced velocity channels maps in the ranges 70--75 km s$^{-1}$, 75--80 km s$^{-1}$, and 80-85 km s$^{-1}$ (Fig.~\ref{fig-channelmaps}). 
The emission mainly peaks in two main condensations, which are clearly seen at 70--75 km s$^{-1}$ (NW\footnote{The 70-75 km s$^{-1}$ map shows another peak farther NW (upper right in the left panel of Fig.~\ref{fig-channelmaps}). However, it falls outside the maximum beam of the single-dish observations studied in this work ($\sim$39\asec), and thus is not discussed.} of the position of G+0.693), and 80--85 km s$^{-1}$ (W). Moreover, the condensation at 80-85 km s$^{-1}$ allows us to study emission that is not contaminated by other gas components. The 75--80 km s$^{-1}$ map is a transition zone between the two condensations. 
In the left and central panels of Fig.~\ref{fig-map-2velcomp}, we highlight the contour corresponding to half of the peak value in the two maps that define the condensations. 
We extracted the spectra from the NW (orange) and W (purple) condensations, which are shown in the rightmost panel of Fig.~\ref{fig-map-2velcomp}. The spectra towards these two condensations peak at velocities that are very similar to the velocities of the narrow velocity components C2 and C3 identified in the previous spectral analysis (Sect.~\ref{res-LTE}). Therefore, we confirm that these two condensations correspond to C2 and C3.
Table \ref{table-coordinates-nonLTE} lists the spatial coordinates of the three velocity components, where C1 is defined from the pointing of single-dish observations, and C2 and C3 are defined from their peak intensities in the velocity maps described above.

From the condensations defined as above, we derive a deconvolved source size\footnote{It was evaluated from the source size of the condensations, $\theta_{\rm c}$, defined as explained above, deconvolved for the beam size: $\sqrt{(\theta_{\rm c})^-(\theta_{\rm beam})^2}$.}, the diameter, of 9\asec\;for the C2 component (corresponding to 0.36 pc at the distance of the source, 8.34$\pm$0.16 kpc \citealt{reid2014}) and of 7.6\asec\;for component C3 (corresponding to 0.3 pc). C2 and C3 might be associated with single condensations with a velocity gradient, or alternatively, with clusters of different condensations. In Appendix \ref{app-spectra-P1-P2-P3}, we show an example of spectra that were extracted from single pixels (i.e. from the SMA beam size), where it is clear that the C2 and C3 line profiles shown in Fig~\ref{fig-map-2velcomp} are composed of even narrower (FWHM$\sim$1.3--4.9 km s$^{-1}$) lines. More sensitive observations with a higher angular and spectral resolution are needed to distinguish these components and clarify their origin (see also the discussion in Sect.~\ref{sec-evolstage}).

To better characterise these condensations, we also produced position-velocity (pv) diagrams in the directions shown in Fig.~\ref{fig-map-2velcomp}, where the zeros are defined at the centre of the condensations. The pv diagrams associated with C2 and C3 are shown in the left and right panels of Fig.~\ref{fig-pv-comp2-comp3}, respectively. C2 clearly peaks at the centre of the condensation at a velocity of 74 km s$^{-1}$. Moreover, southwest of C2 (offset of -5\asec), the gas is connected to higher-velocity gas. Furthermore, no high-velocity component (80--85 km s$^{-1}$) gas is present in the north-eastern part of the region (offset of +5\asec). Regarding C3, we found that the emission peaks at 80.5 km s$^{-1}$ towards the north (offset of -5\asec), connected to gas at 77.5 km s$^{-1}$, also in the northern part of this condensation. 
The velocities found for C2 and C3 are consistent with those found from the LTE analysis. 
The pv diagrams clearly show that the gas components are not independent, but they appear to be connected with each other, as already pointed out at larger scales by \citet{zeng2020} as evidence for a cloud-cloud collision.

\subsection{Non-local thermodynamic equilibrium analysis}
\label{res-non-LTE}

\begin{figure}
\centering
\includegraphics[width=20pc]{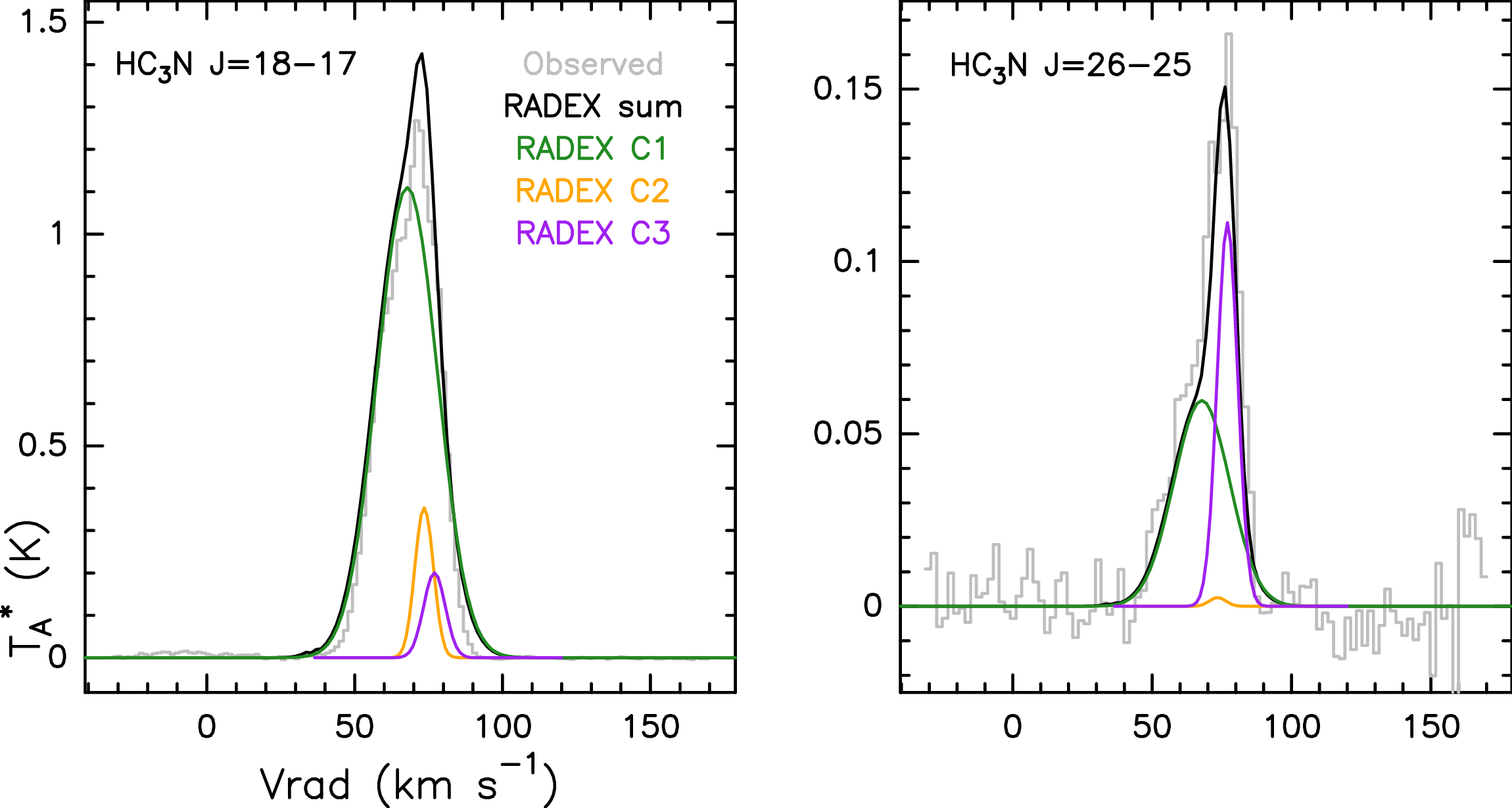}
\caption{Grey histograms show the IRAM 30m observed spectra for the $J$=18--17 and 26--25 transitions (\textit{left} and \textit{righ}t panel, respectively). The solid green, orange, and purple lines are the best non-LTE models to components C1, C2, and C3, respectively, found as described in Sect.~\ref{res-non-LTE}. The solid black lines are the sum of the three components, which match the observed spectra well.}
\label{fig-radex}
\end{figure}

In this section, we present a radiative transfer analysis that we performed for two lines observed with the IRAM 30m, $J$=18--17 and 26--25, which represent the low-$J$ and high-$J$ regimes well and are those in which the three velocity components are better seen. The $J$=18--17 transition is the low-$J$ transition for which component C2 is more prominent, and it is less contaminated by the broad component C1 (which dominates at lower energies). We note that the 19-18 transition was not selected because the spectral resolution is worse. For the high-$J$ group, we used the $J$=26-25 transition because it has the highest energy, where component C3 is better seen, and because it provides a wider range of $E_{\rm up}$.

Components C2 and C3 are affected by beam dilution, beam efficiency corrections, and by the fact that their emission is not centred on the central position of the single-dish observations (C1; see Fig.~\ref{fig-channelmaps}). The last was taken into account through the parameter $s_{\rm shift}$ . $s_{\rm shift}$ was derived from the interferometric cube smoothed with \texttt{MADCUBA} to the IRAM 30m resolutions (15\asec\;and 11\asec\;for the two transitions). For every smoothed cube, we derived the ratio of the emission peak at the velocity of each component (73.5 and 77 km s$^{-1}$ for C2 and C3, respectively), and the emission peak at the same velocity within the single-dish beam centred on C1. For component C2, we derived $s_{\rm shift}$ of 1.18 and 1.35 for $J$=18--17 and 26--25, respectively. For component C3, we derived $s_{\rm shift}$ of 1.28 and 1.44 for $J$=18--17 and 26--25, respectively. Moreover, their observed emission with the single-dish should be converted into the main-beam brightness temperature, $T_{\rm MB}$.

When we take $s_{\rm shift}$, the beam efficiencies, and the beam dilution factors into account (using the source sizes derived above), the $T_{\rm peak}$ to be compared with the radiative transfer simulated spectra are 1.39 K and 0.03 K for the two transitions for component C2, and 0.68 K and 0.28 K for the two transitions for component C3. For component C1, which is extended, we can directly compare with the observed peaks, which are 0.85 K and 0.10 K for the two transitions.

We calculated the intensities using the non-LTE molecular radiative transfer model \texttt{RADEX} (\citealt{vandertak2007}), considering the three gas components separately. 
Table \ref{table-grid} shows the grids of the H$_{2}$ density, $n_{\rm H_{2}}$, and the kinetic temperature, $T_{\rm kin}$, models that we used for the three components. 
In particular, for C1, the density and temperature were derived in previous works (see e.g. \citealt{zeng2018,zeng2020,colzi2022a}) and are about 10$^{4}$ cm$^{-3}$ and $>$100 K, respectively. Before we defined this final grid of models shown in Table \ref{table-grid}, we also confirmed whether higher densities or lower temperatures might explain the observed lines, but without success. For C2, we applied a larger grid of possible density values than for C1, because it was not known. Regarding the temperature, we used a grid with values lower than those used for C1 because we found it to be colder by \citet{colzi2022a}. Finally, for C3, we applied the same grid of density models as for C2. Regarding the temperature, since this component presents a $T_{\rm ex}$ of 64 K (see e.g. Table \ref{table-result}) and $T_{\rm kin}$ cannot be lower than $T_{\rm ex}$ in LTE emission, we restricted the grid we used for C2 starting from a value of 50 K. 
Then, for each component, we used the column density, radial velocity, and line width obtained from the LTE fitting procedure (see Table \ref{table-result}). In particular, for C1 and C2, we used the column densities obtained from the low-$J$ transitions, where these components are more dominant.
For C3, we used the column density obtained from the high-$J$ transitions, because this component mainly appears in high-energy transitions (Fig.~\ref{fig-hc3n-fit}).

From the \texttt{RADEX} outputs obtained with the Python-based script described in Appendix \ref{python-script}, we generated non-LTE synthetic spectra with \texttt{MADCUBA} using the new tool \texttt{Import SPECTRA RADEX FILE} (see Appendix \ref{sec:app-radex} for further information), to compare them with the observed spectra. 
The resulting best \texttt{RADEX} modelled spectra have confidence levels\footnote{This is defined as (Peak$_{\rm RADEX}$-Peak$_{\rm observed}$)/Peak$_{\rm total}$, where Peak$_{\rm RADEX}$ is the peak of the spectra simulated with \texttt{RADEX}, Peak$_{\rm observed}$ is the peak of the observed line component, and Peak$_{\rm total}$ is the total peak of the observed line.} lower than 10\%, 15\%, and 20\% for components C3, C2, and C1, respectively.

The best models found for both transitions are H$_{2}$ density, $n_{\rm H_{2}}$, of 2$\times$10$^{4}$ cm$^{-3}$ and $T_{\rm kin}$=140 K for the broad component C1, $n_{\rm H_{2}}$= 5$\times$10$^{4}$ cm$^{-3}$ and $T_{\rm kin}$=30 K for the narrow component C2, and $n_{\rm H_{2}}$= 4$\times$10$^{5}$~cm$^{-3}$ and $T_{\rm kin}$=80 K for the narrow component C3 (Table \ref{table-coordinates-nonLTE}). The comparison between the observed lines and the non-LTE best model profiles, together with the total emission, is shown in Fig.~\ref{fig-radex}. 
The simulated non-LTE profiles, with the narrow components again converted into $T_{\rm A}^{*}$, reproduce the observed spectra well. This confirms that the column densities derived with the LTE analysis for the different groups of transitions are good approximations for the non-LTE modelling. A similar detailed analysis for different molecular species towards G+0.693 will be presented in a forthcoming paper (Sanz-Novo et al. in prep).

Our analysis has highlighted the presence of distinct gas components in the G+0.693 molecular cloud, with different physical conditions. The diffuse gas observed in component C1 presents a high kinetic temperature that is typical of CMZ molecular clouds (\citealt{krieger2017}). Embedded in the diffuse gas, component C2 shows a higher density and a much lower kinetic temperature, and finally, component C3 shows a density enhancement of more than one order of magnitude with respect to the diffuse gas and also a lower kinetic temperature.
We also found that the difference between the $T_{\rm kin}$ derived from the non-LTE modelling and the $T_{\rm ex}$ derived from LTE decreases when the density of the condensation increases.
$T_{\rm ex}$ derived for component C1 is much lower (15--30 K) than the derived $T_{\rm kin}$ of 140 K (Table \ref{table-coordinates-nonLTE}). This is also similar to what has been derived for other molecular species considering component C1 alone (e.g. \citealt{rivilla2021a,rivilla2022a,colzi2022a}), which are sub-thermally excited due to the relatively low density of C1. However, for components C2 and C3, the difference between the two temperatures decreases ($T_{\rm ex}$=15 K and $T_{\rm kin}$=30 K in C2; $T_{\rm ex}$=65 K and $T_{\rm kin}$=80 K in C3), in agreement with their higher densities, and this indicates that they are closer to LTE. Finally, component C2 presents the lowest $T_{\rm kin}$, also similar to the CMZ dust temperature ($\sim$20 K; \citealt{rodriguez-fernandez2004}), which means that gas and dust are in equilibrium in this source. We discuss these findings and their possible implications in the next section.

\section{Discussion}
\label{discussion}

In this section, we discuss the nature and evolutionary state of the high-density condensations in the context of the cloud-cloud collision framework previously proposed for this region (e.g. \citealt{zeng2020}).

\subsection{Origin of the compact condensations in G+0.693: Cloud-cloud collision}
\label{sec-originevolution}

\begin{figure*}
\centering
\includegraphics[width=42pc]{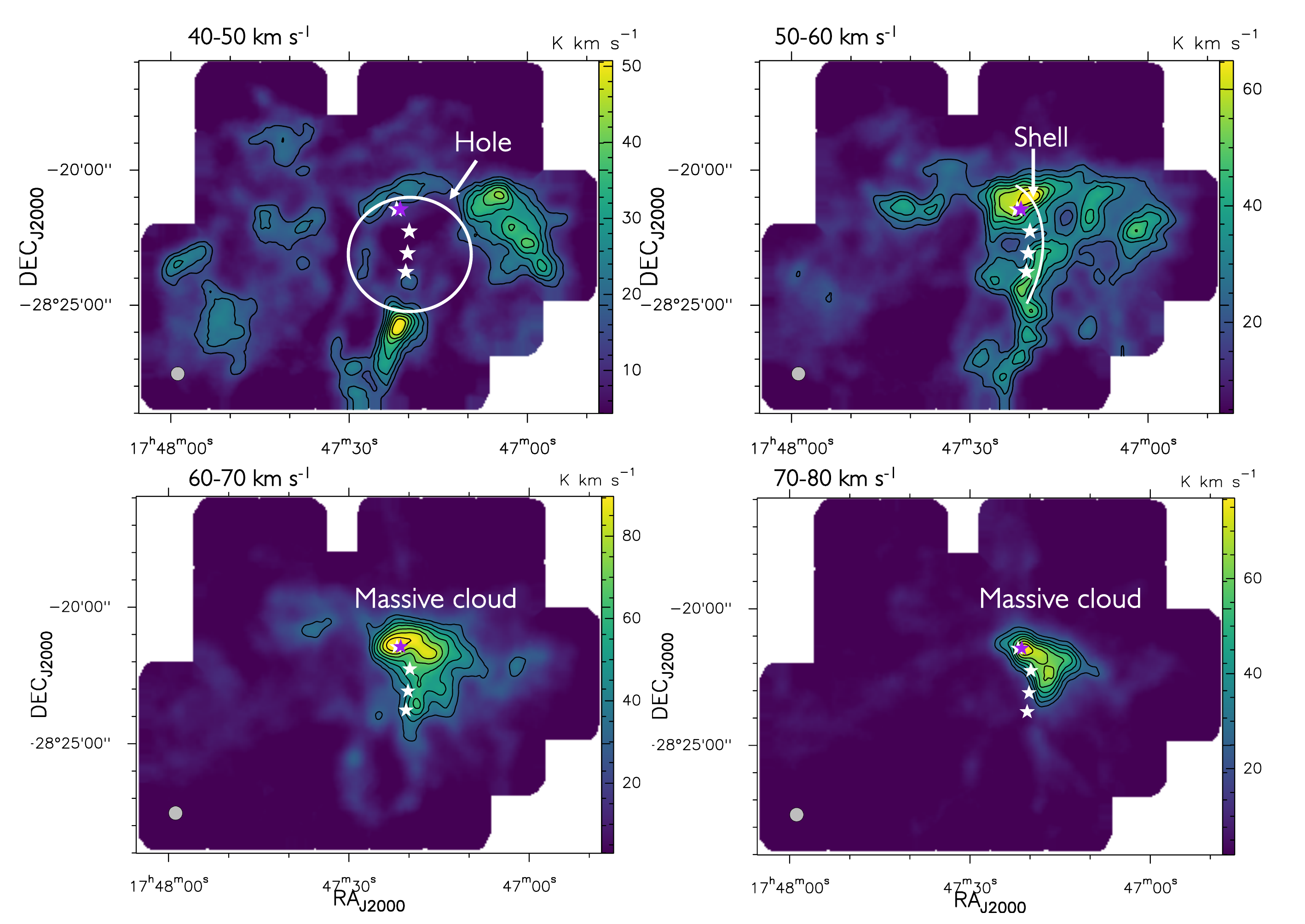}
\caption{HNCO (4$_{0,4}$--3$_{0,3}$) moment 0 maps obtained in the velocity ranges 40--50 km s$^{-1}$ (\textit{top left panel}, the hole), 50--60 km s$^{-1}$ (\textit{top right panel}), 60--70 km s$^{-1}$ (\textit{bottom left panel}, the shell), and 70--80 km s$^{-1}$ (\textit{bottom right panel}, the clump). The contour levels are 0.1 steps from the maximum of the maps (50, 65, 90, and 77 K km s$^{-1}$, respectively). The synthesised beam is the ellipse indicated in the lower left corner (29\asec.5$\times$29\asec.5). The white stars are Sgr B2(S), (M), (N), and G+0.693 from the south to the north, and the purple star is the C3 condensation.}
\label{fig-hnco1}
\end{figure*}

\begin{figure*}
\centering
\includegraphics[width=43pc]{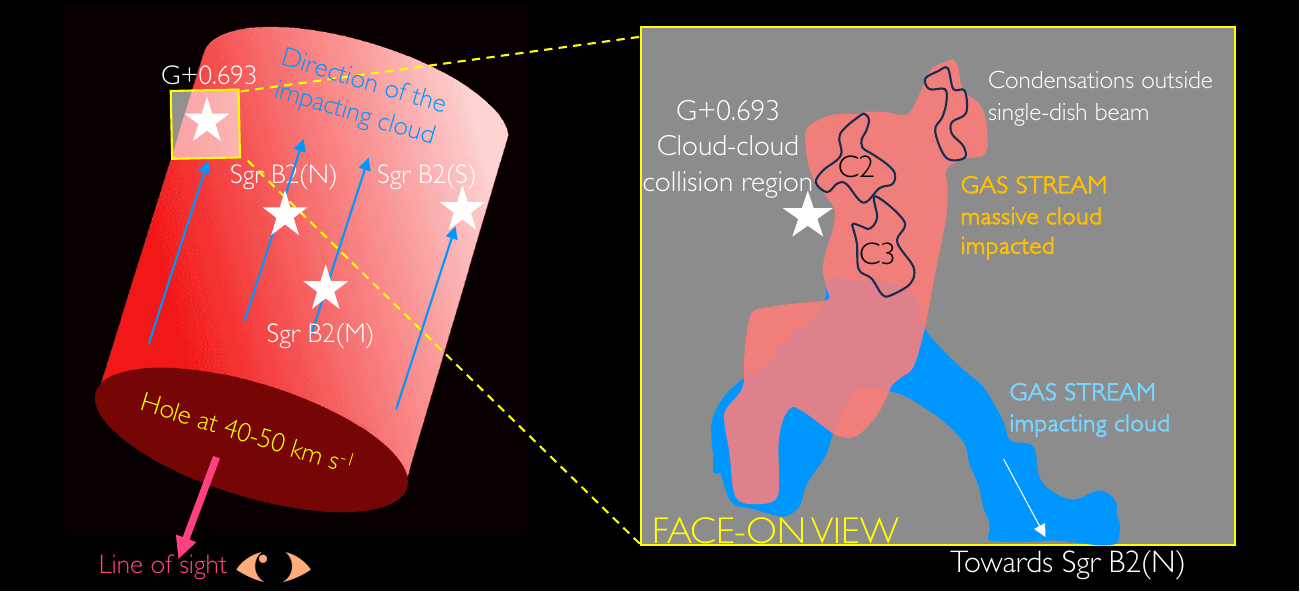}
\caption{Sketch of the Sgr B2 region formation scenario. \textit{Left panel}: In this scenario, Sgr B2(M) is firstly formed during the impact of the small cloud, followed by Sgr B2(N), Sgr B2(S), and G+0.693. These four sources are indicated with white stars. The red cylinder represents the large impacted massive cloud (with velocities of 60--80 km s$^{-1}$), and the blue arrows indicate the possible direction of the smaller impacting cloud, which creates the hole at 40--50 km s$^{-1}$. The transparent white rectangle represents the region that is zoomed-in the right panel. \textit{Right panel}: Face-on view along the line of sight of G+0.693. The blue- and red-shifted regions are drawn from the blue- (50--60 km s$^{-1}$) and red-shifted (70--90 km~s$^{-1}$) CH$_{3}$OH (4--3) integrated intensity by \citet{zeng2020}. The C2 and C3 condensations are those defined in Fig.~\ref{fig-pv-comp2-comp3}. The north-west region represents a group of condensations seen in the left panel of Fig.~\ref{fig-channelmaps}, but which mainly falls outside the single-dish beams of the observations used in this work (see Table \ref{table-HC3Ntransitions}). The white star shows the central position of the single-dish observations and probably is the centre of the cloud-cloud collision between the two gas streams. The red gas stream, border of the cylinder, represents the massive cloud that is impacted and tracked by the smaller cloud, i.e. the blue gas stream.}
\label{fig-sketch1}
\end{figure*}

In Sect.~\ref{res-non-LTE} we derived the H$_{2}$ densities and $T_{\rm kin}$ for the three gas components (see Table~\ref{table-coordinates-nonLTE}), finding that the gas becomes denser and colder in the condensations C2 and C3 with respect to the less-dense and warmer gas (C1). 
This is consistent with a shock origin for C2 and C3, which are both in a post-shock evolutionary stage. It is already known that G+0.693 presents clear signatures of a cloud-cloud collision, together with small-scale structures ($\sim$0.3 pc). \citet{zeng2020} unveiled that two main large-scale molecular components at 57 km s$^{-1}$ and 78 km s$^{-1}$ are connected in space and velocity, consistent with the kinematic features revealed by \citet{hasegawa1994} towards Sgr B2. They found a hole at 50 km s$^{-1}$ (a blue-shifted cloud) that is spatially related with a shell and a massive cloud seen in emission (a molecular cloud and a maser) over the velocity ranges 60--70 and 70--80 km s$^{-1}$ (a red-shifted cloud), respectively. The same structures can be seen in the velocity-integrated maps of the well-known shock tracer HNCO (4$_{0,4}$--3$_{0,3}$ transition)\footnote{IRAM 30m mosaicking observations from the project 133-19 (PI: Rivilla), which will be presented in a forthcoming paper.} shown in Fig.~\ref{fig-hnco1}. These maps further support the scenario that has been hypothesised by other authors (\citealt{sato2000,armijos-abendano2020,enokiya2022}), in which during a large-scale cloud-cloud collision, a dense cloud is compressed and hollowed out by a smaller cloud that punches the dense, massive cloud. G+0.693 is located in the interface region of the hole and the massive cloud with little emission in the 57 km s$^{-1}$ component and with a peak in the 78 km s$^{-1}$ component. As shown in Fig.~\ref{fig-hnco1}, the massive cloud hosts a ridge of recent massive star formation (Sgr B2(S), Sgr B2(M), and Sgr B2(N), which are indicated as white stars), and G+0.693 is also located in the northern part of this star formation ridge. The left panel in fig.~\ref{fig-sketch1} sketches the shock evolution in the cloud-cloud collision scenario that likely triggered the sequential formation. In this scenario, the small cloud first shocked the Sgr B2(M) region, which seems to be in a later stage of evolution (e.g. \citealt{devicente2000,schmiedeke2016}), and then, it impacted the Sgr B2(N) region, in which multiple protostars are forming (e.g. \citealt{belloche2016}), and finally, it produced the Sgr B2(S) star-forming region (e.g. \citealt{jeff2024}).  

G+0.693 is located in the northern region of the ridge and, since it is symmetrically located with respect to Sgr B2(S) (see Fig.~\ref{fig-sketch1}), it is an excellent candidate for hosting the next massive star formation event. \citet{henshaw2016} suggested a cone-like structure of Sgr B2 in a HNCO position-position-velocity diagram, with G+0.693 located close to the tip of the conical structure, as expected for a cloud-cloud collision (see e.g. \citet{zeng2020} for a detailed description).
The right panel in Fig.~\ref{fig-sketch1} shows the small-scale structure of the shocked gas in the G+0.693. The blue- and red-shifted CH$_{3}$OH emission from \citet{zeng2020} would correspond to the two gas streams generated in the cloud-cloud collision. The blue part represents the remains of the impacting cloud, and the red part shows the post-shock condensations discussed in this paper. 
The shocks towards G+0.693, which have increased the density by more than a factor of 5 (from the diffuse C1 up to the densest C3), are very likely the origin of the observed condensations (Fig.~\ref{fig-sketch1}). Cloud-cloud collisions have also been found to trigger star formation in many high-mass star-forming regions throughout the Milky Way, and the kinematic properties described above are typical for the prediction of cloud-cloud collisions in hydrodynamic simulations (see e.g. \citealt{habe1992,armijos-abendano2020,fukui2021}).

\subsection{Possible evolution of the condensations}
\label{sec-evolstage}

\begin{table*}
\begin{center}
\caption{\label{table-masses} Physical properties of the two condensations towards G+0.693.}
\begin{tabular}{lccccccc}
\hline
Component & Size\tablefootmark{a} & Mass & $M_{\rm vir}$ & $\alpha$ & $M_{\rm J}$ & $\lambda_{\rm J}$& $M_{\rm J}^{\rm turb}$ \\
& (pc) & (M$_{\odot}$) & (M$_{\odot}$) & &(M$_{\odot}$) & (pc)&(M$_{\odot}$) \\
\hline
C2  & 0.36& 68 & 2$\times$10$^{3}$ & 29& 5.7 & 0.15 & 5.5$\times$10$^{3}$\\
C3  & 0.3 & 330  &2.5$\times$10$^{3}$ & 7.6 & 8.8 & 0.08 & 3.6$\times$10$^{3}$\\
\hline
C2*\tablefootmark{b} & 0.18 & 8 & 32--451 & 4--56 &5.7 &0.15 &30--1.74$\times$10$^{3}$ \\
C3*\tablefootmark{c} & 0.18 & 71 & 32--451& 0.45--6.3 &8.8 &0.08&10--620 \\
\hline
\normalsize
\end{tabular}
\end{center}
\tablefoot{The third column lists the mass of the gas. Columns 4 and 5 list the virial mass and the virial parameter, respectively. The last three columns are the Jeans mass, the length, and the turbulent mass. The formulas we used are described in Sect.~\ref{sec-evolstage}. \tablefoottext{a}{The size corresponds to the diameter.} \tablefoottext{b}{Assuming a radius that is half of the SMA beam (4.5\asec/2), using C2 physical conditions, and a FWHM range of 1.3--4.9 km s$^{-1}$.} \tablefoottext{c}{Assuming a radius of (4.5\asec/2), using the physical conditions of C3, and a FWHM range of 1.3--4.9 km s$^{-1}$.} 
}
 \end{table*}


It is well known that star formation in the GC is strongly suppressed, but a few sites show recent massive star formation (e.g. \citealt{longmore2013}). The Sgr B2 complex stands out as one of the few regions in which clusters of massive stars form (e.g. \citealt{devicente2000,ginsburg2016}). Three clusters of massive stars have recently formed in Sgr B2 along the high-density ridge that was likely created by a cloud-cloud collision. This type of event is expected to trigger high-mass cluster formation in galaxies (\citealt{fukui2021, rico-villas2022}). In this scenario, the shock compresses the gas in the interface region into a dense thin layer, in which non-linear hydrodynamic processes lead to fragmentation (e.g. \citealt{heitsch2008}), together with a rise in gas densities that is compatible with future star formation (e.g. \citealt{vanloo2007,cosentino2019,cosentino2023}). \citet{vanloo2007} found with a magnetohydrodynamics code that a dense shell can form within a shock and can fragment subsequently. The authors suggested that this might be the primary region for the formation of massive stars. The processes leading to  fragmentation in molecular clouds remain an open problem, especially in the GC, where non-thermal pressure, such as magnetic fields and/or turbulence, can dominate the process. The three sites of massive star cluster formation in Sgr B2 are basically aligned and regularly spaced along the high-density ridge (left panel of Fig.~\ref{fig-sketch1}), suggesting a large-scale filamentary gravitational instability with a spatial scale of $\sim$2 pc. Remarkably, the G+0.693 high-density condensations are located just 2 pc north of Sgr B2(N), following the same trend as found for the other sites of massive star formation, and supporting the idea that the G+0.693 condensations studied in this work represent prestellar stages of massive star clusters.  

Based on the physical and kinematic parameters derived for the C2 and C3 condensations, we discuss their possible evolutionary stage below. 

The cloud that lies between 70 and 85 km s$^{-1}$ (Fig.~\ref{fig-vel-70-85}) presents a filament-like structure, which could be the result of one or multiple shock-compressed gas layers (see Sect.~\ref{sec-originevolution}). The pv diagrams in Fig.~\ref{fig-pv-comp2-comp3} can be interpreted as a shock that first encounters C2, which has already cooled down ($T_{\rm kin}$= 30 K), and that more recently impacted C3, which is still cooling down (with $T_{\rm kin}$= 80 K), compared to the 140 K of the diffuse component C1. The evolution of the temperature during the shock passage described above is expected from the results of C-shock models (e.g.~\citealt{jimenez-serra2008}), with $T_{\rm kin}$ of the neutral gas increasing to $>$1000 K during the shock, and cooling to $<$100 K in the post-shock environment after $\sim$10$^{3}$ years. In this shock interaction, turbulence dissipation is also expected (e.g.~\citealt{lesaffre2013}), which explains the narrower line widths of C2 and C3 with respect to the extended component C1.
\citet{colzi2022a} studied the D/H ratios of HCN, HNC, HCO$^{+}$, and N$_{2}$H$^{+}$ towards G+0.693 using multiple rotational transitions. In addition to the turbulent line component with a FWHM of 20 km s$^{-1}$, they also revealed a less turbulent component with a FWHM of 9 km s$^{-1}$ (narrow component), with higher D/H values. These authors estimated  $T_{\rm kin}$ of 100 K and H$_{2}$ densities of (0.3--3)$\times$10$^{4}$ cm$^{-3}$ for the broad component, and $T_{\rm kin}\leq$30 K and H$_{2}$ densities that increased by at least one order of magnitude, (0.05--1)$\times$10$^{6}$ cm$^{-3}$, for the narrow component. These two components are indeed consistent in velocity, density, and temperature with C1 and C2. 
Component C3 is not present in the deuterated molecules studied by \citet{colzi2022a}, as expected because the molecular transitions studied by these authors have $E_{\rm up}<$25 K, while C3 is mainly traced by higher-energy transitions with $E_{\rm up}>$100 K.

Important physical parameters that should be taken into account in a study of the evolution of the condensations are their mass, $M$, size, and stability. So-called subcritical clouds are unbound, and may expand and dissolve into the diffuse ISM, while supercritical clouds are (marginally) gravitationally bound, and can undergo collapse when perturbed. 
The cloud mass can been calculated as
\begin{equation}
M = \rho\frac{4}{3}\pi\biggl(\frac{R}{2}\biggr)^{3},
\end{equation}
where $\rho$ is the mean mass density derived by taking the mean molecular weight of 2.33 into account for fully molecular gas with a 10\% He:H ratio, and $R$ is the deconvolved source size (the diameter), given in Table \ref{table-result}, in cm. We derived $M$ of $\sim$1.4$\times$10$^{35}$\,g and $\sim$6.5$\times$10$^{35}$\,g for C2 and C3, respectively, which translates into 68 M$_{\odot}$ and 330 M$_{\odot}$ for C2 and C3, respectively (Table \ref{table-masses}).

To investigate the stability of these clouds against collapse, we compared their measured properties with those predicted based on different criteria. 
 In general, a cloud is stable against fragmentation when its mass is lower than the gravitational critical Jeans mass, $M_{\rm J}$, of a non-magnetic isothermal clump,
\begin{equation}
\label{eq-mass-jeans}
M_{\rm J}= 2.47\;{\rm M}_{\odot} \biggl(\frac{T_{\rm kin}}{10\;{\rm K}}\biggr)^{3/2} \biggl(\frac{n_{\rm H_2}}{10^{4}\;{\rm cm}^{-3}}\biggr)^{-1/2},
\end{equation}
where $n_{\rm H_2}$ is the H$_{2}$ density of the cloud (\citealt{stahlerpalla}). We found $M_{\rm J}$ = 5.7 and 8.8 M$_{\odot}$ for C2 and C3, respectively, which are much lower than the masses of the two condensations. Thus, we would expect the condensations to fragment at scales of the Jeans length, which is given by 
\begin{equation}
\label{eq-lambda-jeans}
    \lambda_{\rm J}= \biggl(\frac{\pi\;a_{\rm T}^{2}}{\rho\;G}\biggr)^{1/2}= 0.19\;\text{pc}\; \biggl(\frac{T_{\rm kin}}{10\;{\rm K}}\biggr)^{1/2} \biggl(\frac{n_{\rm H_2}}{10^{4}\;{\rm cm}^{-3}}\biggr)^{-1/2},
\end{equation}
where $a_{\rm T}$ is the sound speed for an isothermal clump, and $G$ is the gravity constant. The corresponding Jeans lengths (see Table \ref{table-masses}) are $\lambda_{\rm J}$ = 0.15 and 0.08 pc for C2 and C3, respectively, which are close to the spatial resolution of the SMA observations of about 0.18 pc. This could suggest unresolved Jeans mass fragments within C2 and C3. 

In the GC, condensations also contain non-thermal pressure that is dominated by turbulence and/or magnetic fields. It is still debated what under these conditions controls the fragmentation process, but it is expected that turbulence or a magnetic field could play an important role (\citealt{fukui2021}). The non-thermal motions in C2 and C3, as measured from the observed line widths, are clearly supersonic, and hence, eq.~\eqref{eq-mass-jeans} can be rewritten as
\begin{equation}
\label{eq-mass-jeans-turb}
M_{\rm J}^{\rm turb}= \biggl(\frac{2.9\;\sigma_{\rm turb}^{3}}{\rho^{1/2}\;G^{3/2}}\biggr),
\end{equation}
considering that the mass is contained in a sphere of diameter $\lambda_{\rm J}$, and $a_{\rm T}$ was replaced by the one-dimension turbulent velocity dispersion, $\sigma_{\rm turb}$, 
\begin{equation}
    \sigma_{\rm turb}^{2} = \sigma_{\rm v}^2 - \frac{kT_{\rm kin}}{\mu_{\rm HC_{3}N}m_{\rm H}},
\end{equation}
where $k$ is the Boltzmann constant, $\mu_{\rm HC_{3}N}$ is the HC$_{3}$N molecular weight, and $m_{\rm H}$ is the mass of a hydrogen atom. The first term, $\sigma_{\rm v}$, is the one-dimension velocity dispersion, FWHM/2.35 assuming Gaussian line profiles (e.g. \citealt{bertoldi1992, kauffmann2013}). The second term is the thermal contribution corresponding to the sound speed of the observed molecule. For HC$_{3}$N, with a molecular weight of 51, this term is lower than 0.1 km s$^{-1}$, and thus, it is negligible with respect to the one-dimension velocity dispersion for the two condensations ($\sim$3--4 km s$^{-1}$). Thus, we can derive the Jeans turbulent mass from the measured line widths, and we obtain $M_{\rm Jeans}^{\rm turb}$, of 5.5$\times$10$^{3}$ M$_{\odot}$ and 3.6$\times$10$^{3}$ M$_{\odot}$ for C2 and C3, respectively (Table \ref{table-masses}).

The observed velocity gradients in the region make unclear to which extent the measured HC$_{3}$N line widths from our single-dish spectra directly translate into the internal pressure. To reduce the contribution of the velocity gradients to the total line width, we used interferometric data with a resolution of $\sim$4.5\asec\;to measure individual line widths for each condensation. As we showed in Sect.~\ref{res-maps} and Appendix \ref{app-spectra-P1-P2-P3}, single pixels (i.e. spectra extracted from a beam size of $\sim$0.2 pc) contain narrower velocity components, as expected (see Fig.~\ref{fig-spectra-P1-P2-P3}).  The fitted line widths range from 1.3 to 4.9 km s$^{-1}$.
Table \ref{table-masses} shows the values of the masses and the range of turbulent Jean masses of the two condensations considering this lower velocity dispersion (indicated as C2* and C3* in the following). While the colder C2* condensation, with a lower mass than $M_{\rm Jeans}^{\rm turb}$, seems to be already stable against further fragmentation or collapse, the warmer and denser C3* condensation seems to be unstable against fragmentation or gravitational collapse. This suggests that turbulence could play an important role in the fragmentation process, as was found in other cloud-forming massive stars in the Galactic disk (e.g. \citealt{wang2011}).

In  Table \ref{table-masses} we also list the virial parameter, $\alpha$, and the virial masses derived assuming no external pressure and no magnetic field as
\begin{equation}
\label{eq-viral}
\alpha = \frac{M_{\rm vir}}{M} \equiv \frac{5\sigma_{\rm v}^{2}(R/2)}{GM},
\end{equation}
where $M_{\rm vir}$ = 5$\sigma_{\rm v}^{2}(R/2)$/$G$ is the virial mass. The derived  $\alpha$ value ranges from 0.45 to 56 (Table \ref{table-masses}). To minimise the effect of the velocity gradients in $\alpha$, we focused on the condensations C2* and C3*, and we took the narrower single-pixel velocity components into account. Clouds with $\alpha\gg$ 2 contain enough kinetic energy to expand and dissolve into the surrounding environment, meaning that the gravity of the cloud might not support it against expansion. Conversely, if $\alpha\ll$ 2, there is no sufficient internal pressure, and the cloud will be unstable and will collapse (\citealt{kauffmann2013}). 
For C2*, $\alpha$ reaches values of 56, indicating that it might still expand, unless it is confined by external pressure (e.g. \citealt{bertoldi1992}). For C3*, the $\alpha$ values of 0.45--6.3 would indicate that this condensation could be unstable and could collapse. This is also consistent with the turbulent analysis above.  
We note than this basic virial analysis strongly depends on the angular resolution of the observations. Even at scales of 0.2 pc, the velocity gradients in the cloud-cloud interface can contribute much to the measured line width.

Our data suggest the shock-triggered formation of a prestellar condensation in the CMZ that might be on the verge of collapse. High angular resolution and sensitive observations of multiple transitions of shock and density tracers, for example SiO, HNCO, and HC$_{3}$N, are needed to study the shock structures in detail and to characterise its kinematic and physical properties to firmly establish the evolution of the high-density condensations found in the G+0.693 region.

\section{Conclusions}
\label{conclusions}

We have suggested the presence of a prestellar condensation in the CMZ. We studied the temperature and density structure of G+0.693, a molecular cloud located in the CMZ in the northeast part of the Sgr B2 region, with a multiple transition analysis of HC$_{3}$N from single-dish observations, coupled with spatially resolved images of the same molecule. The main results and conclusions of our study are summarised below.
\begin{itemize}
\item[1.] The LTE analysis of 18 HC$_{3}$N rotational transitions, $J$, obtained with the Yebes 40m, IRAM 30m, and APEX radio telescopes, allowed us to first distinguish a different behaviour for low-$J$ (up to 19--18) and high-$J$ ($J>$19--18, with $E_{\rm up}>$ 100 K) transitions due to non-LTE effects. Moreover, the spectral shape of these transitions showed a first broad (FWHM$\sim$23 km s$^{-1}$) component (C1) and a second narrow peak at higher velocities, which in turn is explained by two components (C2 and C3) with a FWHM of 7.2 and 8.8 km s$^{-1}$, respectively. The presence of narrow components is consistent with what was found by \citet{colzi2022a} from isotopologues of HCN, HNC, HCO$^{+}$, and N$_{2}$H$^{+}$. The final column densities obtained are $N$= (6.54$\pm$0.07)$\times$10$^{14}$ cm$^{-2}$, (9$\pm$3)$\times$10$^{14}$ cm$^{-2}$, and (3.6$\pm$0.7)$\times$10$^{13}$ cm$^{-2}$, for C1, C2, and C3, respectively, which is consistent within a factor 3$\sigma$ with the values obtained with the rotational diagram analysis. 
\item[2.] We performed a radiative transfer analysis for two representative lines of the low-$J$ (18--17) and high-$J$ (26--25) regimes. We found H$_{2}$ densities of 2$\times$10$^{4}$ cm$^{-3}$, 5$\times$10$^{4}$ cm$^{-3}$, and 4$\times$10$^{5}$ cm$^{-3}$, and kinetic temperatures of 140 K, 30 K, and 80 K for C1, C2, and C3, respectively. Moreover, higher angular resolution maps obtained with the SMA and APEX for the HC$_{3}$N(24-23) transition clearly show the spatial extent of C2 and C3 in the velocity ranges 70--75 and 80--85 km s$^{-1}$, which is consistent with the velocities found from the LTE analysis. We found that component C1 is extended (>1 pc), and we found deconvolved source sizes of 9\asec\;(0.36 pc) and 7.6\asec\;(0.3 pc) for C2 and C3, respectively. These results highlight high-density and colder condensations (C2 and C3) that are embedded in a more diffuse and warm gas (C1) in the G+0.693 molecular cloud.
\item[3.] The physical properties we found for C2 and C3 are consistent with a shock origin of the two condensations, where the two sources are in a post-shock evolutionary stage. This is in line with the fact that G+0.693 has been proposed to be in the centre of a cloud-cloud collision. Previous works about the Sgr B2 region reported a hole at 40--50 km s$^{-1}$ that is spatially related with a shell at 50--60 km s$^{-1}$ and a massive cloud at 60--80 km s$^{-1}$, which was also confirmed by new HNCO (4--3) IRAM 30m maps. We propose the following scenario, sketched in Fig.~\ref{fig-sketch1}: A small cloud shocked the Sgr B2 region, firstly forming Sgr B2(M). It then impacted Sgr B2(N), and finally, Sgr B2(S) and G+0.693. Zooming into G+0.693, the blue-shifted tracked molecular emission represents the remains of the impacting cloud, and the red-shifted emission studied in this work is the post-shocked emission in which the condensations formed.
\item[4.] In the shock scenario, C2 is probably impacted first and is already cold, while C3 was recently impacted and is still cooling down. A pixel-by-pixel inspection showed that they are composed of even narrower lines (FWHM of 1.3--4.9 km s$^{-1}$), indicating that the internal turbulence of the condensations is caused by internal velocity gradients. This means that the actual line widths may be narrower than observed at the limited angular resolution of the SMA images. We investigated the possible evolutionary stage of the condensations through the analysis of their Jeans masses and virial parameters. Taking their internal turbulence into account, we found that G+0.693 has undergone the process of internal fragmentation, C2 is stable against further fragmentation and might still expand, and C3 is not stable and could collapse.
\end{itemize}

\begin{acknowledgements}
We thank the anonymous referee for the careful reading of the article and the useful comments. We are very grateful to the Yebes 40m, IRAM 30m, and GBT telescope staff for their precious help during the different observing runs and to the APEX staff for conducting the observations. The 40m radio telescope at Yebes Observatory is operated by the Spanish Geographic Institute (IGN; Ministerio de Transportes, Movilidad y Agenda Urbana). IRAM is supported by INSU/CNRS (France), MPG (Germany) and IGN (Spain). The Green Bank Observatory is a facility of the National Science Foundation operated under cooperative agreement by Associated Universities, Inc. APEX is supported by the
Max-Planck-Institut fuer Radioastronomie. 
L.C., J.M-P, I.J-S., and V.M.R. acknowledge support from grant No. PID2019-105552RB-C41 by the Spanish Ministry of Science, Innovation and Universities/State Agency of Research MICIU/AEI/10.13039/501100011033. Moreover, L.C., I.J-S., V.M.R., and M.S-N. acknowledge support from grant No. PID2022-136814NB-I00 by MICIU/AEI/10.13039/501100011033 and by ERDF, UE. V.M.R. also acknowledges support from the grant number RYC2020-029387-I funded by MICIU/AEI/10.13039/501100011033 and by "ESF, Investing in your future", and from the Consejo Superior de Investigaciones Cient{\'i}ficas (CSIC) and the Centro de Astrobiolog{\'i}a (CAB) through the project 20225AT015 (Proyectos intramurales especiales del CSIC); and from the grant CNS2023-144464 funded by MICIU/AEI/10.13039/501100011033 and by “European Union NextGenerationEU/PRTR”. M.S-N. also acknowledges a Juan de la Cierva Postdoctoral Fellowship, project JDC2022-048934-I, funded by MCIN/AEI/10.13039/501100011033 and by the European Union “NextGenerationEU/PRTR”. X.L. acknowledges support from the National Key R\&D Program of China (No.\ 2022YFA1603101), the Natural Science Foundation of Shanghai (No.\ 23ZR1482100), the National Natural Science Foundation of China (NSFC) through grant Nos. 12273090 \& 12322305, and the Chinese Academy of Sciences (CAS) `Light of West China' Program (No.\ xbzgzdsys-202212). Q. Z. acknowledges the support of National Science Foundation under award No. 2206512.
\end{acknowledgements}

%
%

\bibliographystyle{aa} 
 \bibliography{bibliography} 

\begin{appendix} 

\section{Additional transitions}
\label{app-other-transitions}
\begin{figure}[h!]
\centering
\includegraphics[width=22pc]{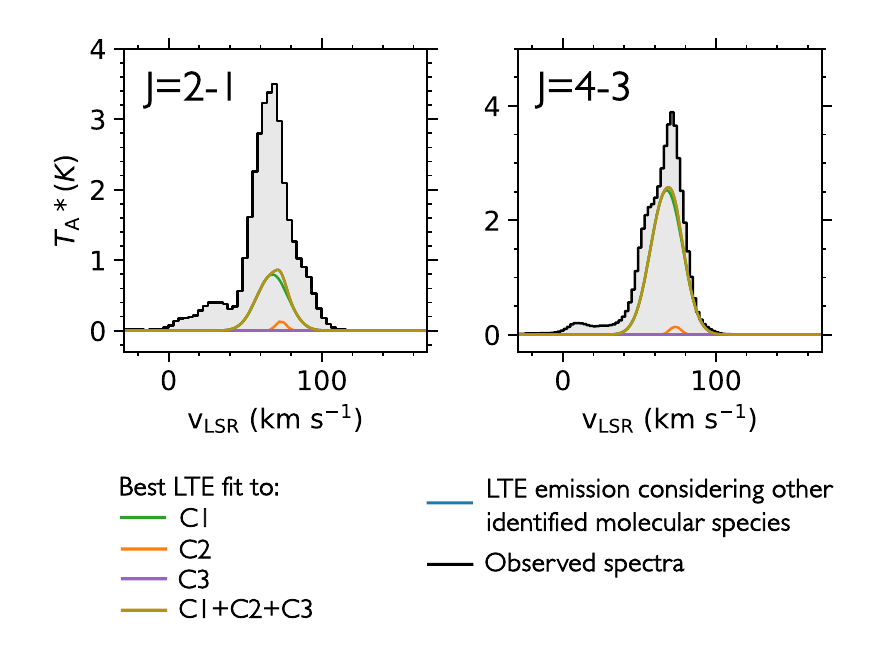}
\caption{Same as Fig.~\ref{fig-hc3n-fit} but for the transitions $J$=2--1 (\textit{left panel}) and 4--3 (\textit{right panel}). The best LTE fit obtained for the "low-$J$" transitions is used in the plot.}
\label{fig-hc3n-nofit}
\end{figure}

In this appendix the detected rotational transitions not used for the analysis described in Sect.~\ref{results} are shown (Fig.~\ref{fig-hc3n-nofit}). The 2--1 transition has been observed with the GBT radio telescope with a beam of 42\asec, while the 4--3 transition has been observed with the Yebes 40m radio telescope resulting in beam of 48\asec. Their line intensities are higher than those predicted by the LTE fit for the "low-$J$" transitions (see Sect.~\ref{res-LTE}), as well as the line profiles differs from the other transitions (Fig.~\ref{fig-hc3n-fit}). This suggests the presence of additional gas component not studied in this work.


\section{Source size of component C3}
\label{app-sourcesize}

\begin{figure*}
\centering
\includegraphics[width=30pc]{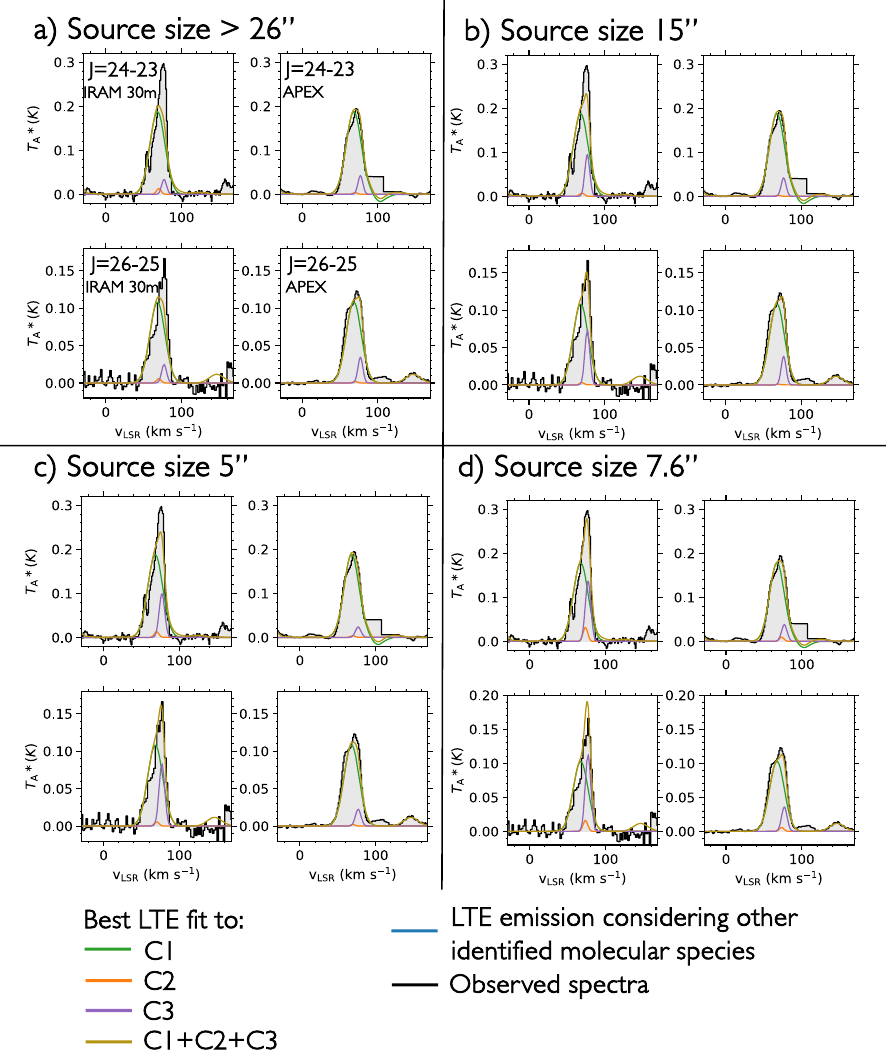}
\caption{Same as Fig.~\ref{fig-hc3n-fit} but for the transitions $J$=24--23 and 26--25, observed both with the IRAM 30m and APEX. Each of the four panel represent the best LTE fitted spectra obtained changing the source size for the C3 (purple) component. A source size smaller than 10\asec\;is needed to explain the higher velocity peaks seen with the IRAM 30m, i.e. with a lower angular resolution than with APEX.}
\label{fig-hc3n30mapex}
\end{figure*}

We show here how assuming different source sizes change the LTE simulated spectra of the HC$_{3}$N(24--23) and (26--25), observed with both the IRAM 30m and APEX radio telescopes.
It is clear from Fig.~\ref{fig-hc3n30mapex} that the same transition presents a different line profile at higher velocities if observed with the IRAM 30m ($\sim$11\asec\,at 230 GHz) or with APEX ($\sim$26\asec\,at 230 GHz). The peak at $\sim$77 km $^{-1}$ observed with the IRAM 30m and not with APEX indicates that the emission of the C3 narrow component is more compact than 26\asec. 

To estimate the best source size we have assumed different source sizes, and then beam dilution factors, for component C3, of 5\asec, 15\asec, and extended emission, as well as the source size estimated from the maps in Sect.~\ref{res-maps} of 7.6\asec. Then, we have performed the fit procedure to the high-$J$ spectra as explained in Sect.~\ref{res-LTE}, keeping fixed all the parameters, except the total column density, to those found in Table \ref{table-result}. As shown in Fig.~\ref{fig-hc3n30mapex} the source size of 7.6\asec is the best to reproduce the observed spectral profiles.

\clearpage
\section{Additional velocity channel maps and spectra}
\subsection{The HC$_{3}$N emission in 50--70 km s$^{-1}$}
\label{app-vel-50-60}

In this appendix we show the velocity channel map in the range 50--70 km s$^{-1}$, not shown in Fig.~\ref{fig-channelmaps}. Most of this emission peaks towards the southern-west. This is probably another compact gas component, which is completely blended by the broad extended gas component in the single-dish spectra (Fig.~\ref{fig-hc3n-fit}). It is also consistent with the blue-shifted emission of the possible cloud-cloud collision present towards G+0.693 studied by \citet{zeng2020}.

\begin{figure}[h!]
\centering
\includegraphics[width=20pc]{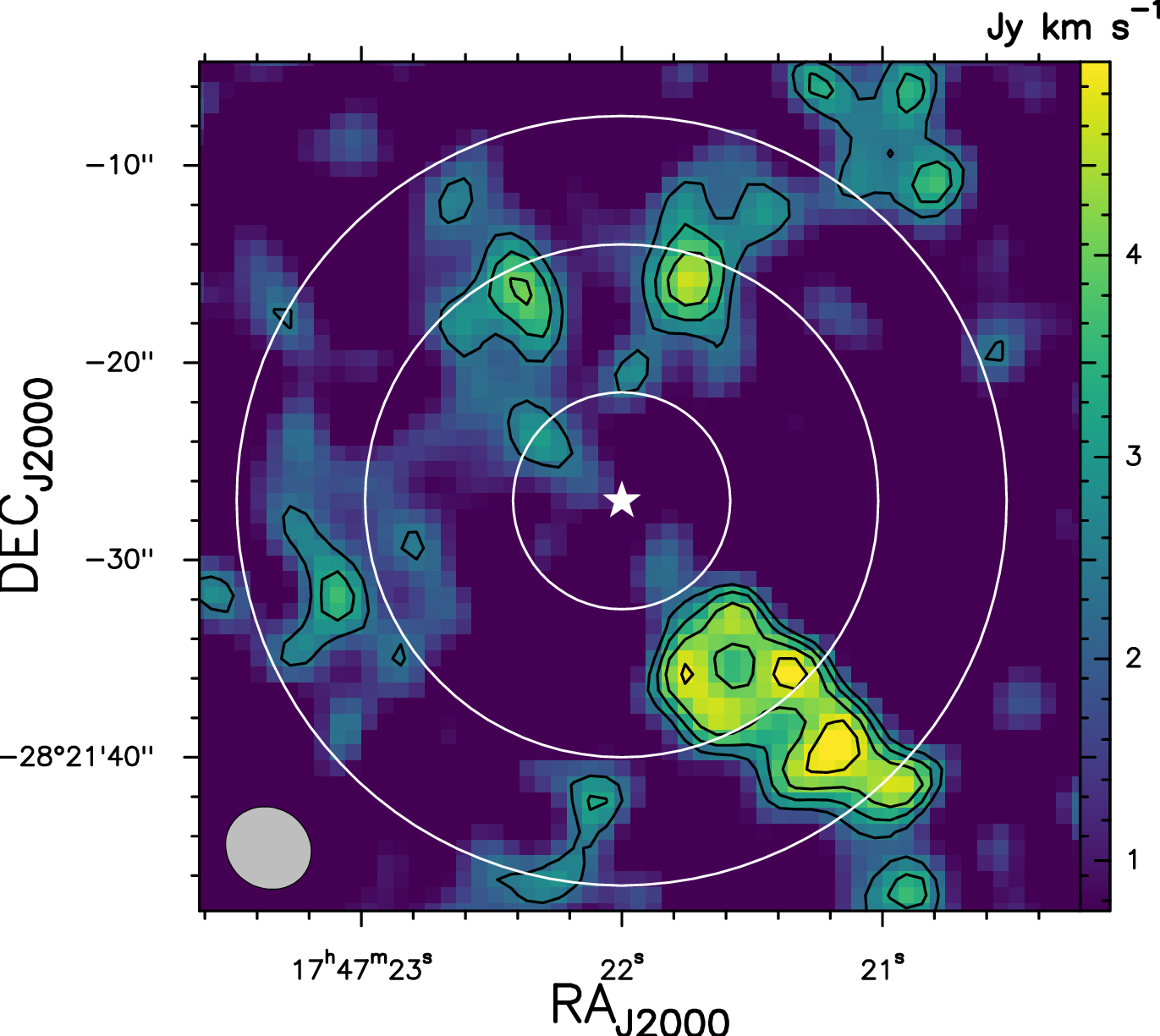}
\caption{Same as Fig.~\ref{fig-hc3n_24_23_all} but in the velocity range 50-70 km s$^{-1}$. The 1$\sigma$ rms is 0.750 Jy km s$^{-1}$.}
\label{fig-vel-50-60}
\end{figure}

\subsection{The HC$_{3}$N emission in 70--85 km s$^{-1}$}
\label{app-vel-70-85}
In this appendix we show the velocity channel map in the range 70--85 km s$^{-1}$. These velocities are those in which the narrow components are defined and presents a filament-like structure as discussed in Sect.~\ref{discussion}.

\begin{figure}[h!]
\centering
\includegraphics[width=20pc]{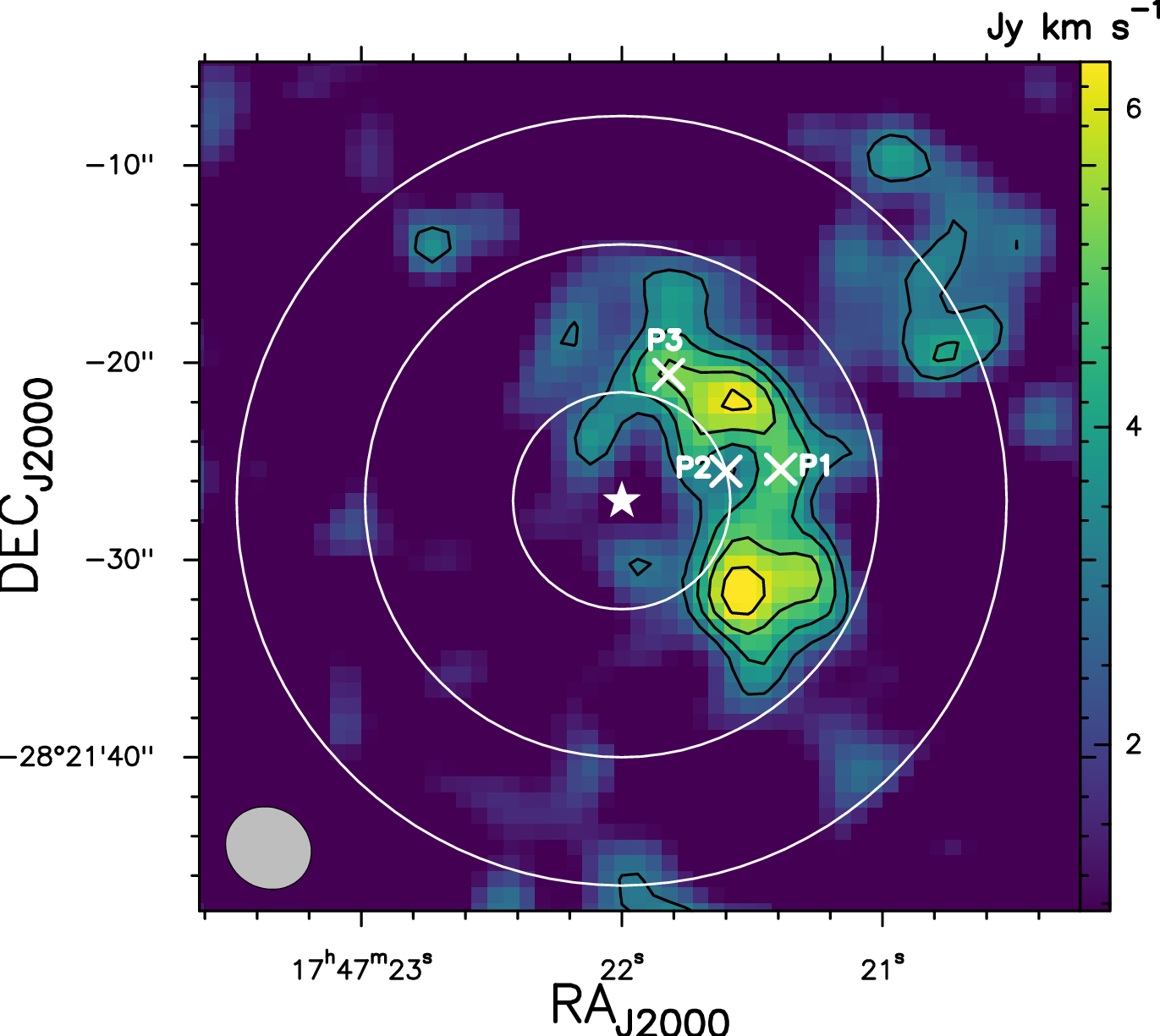}
\caption{Same as Fig.~\ref{fig-hc3n_24_23_all} but in the velocity range 70-85 km s$^{-1}$. The 1$\sigma$ rms is 0.954 Jy km s$^{-1}$. White crosses are P1, P2, and P3 positions where single pixel spectra has been extracted (see Sect.~\ref{app-spectra-P1-P2-P3}).}
\label{fig-vel-70-85}
\end{figure}

\subsection{Spectra from single pixels}
\label{app-spectra-P1-P2-P3}
\begin{figure*}[h!]
\centering
\includegraphics[width=40pc]{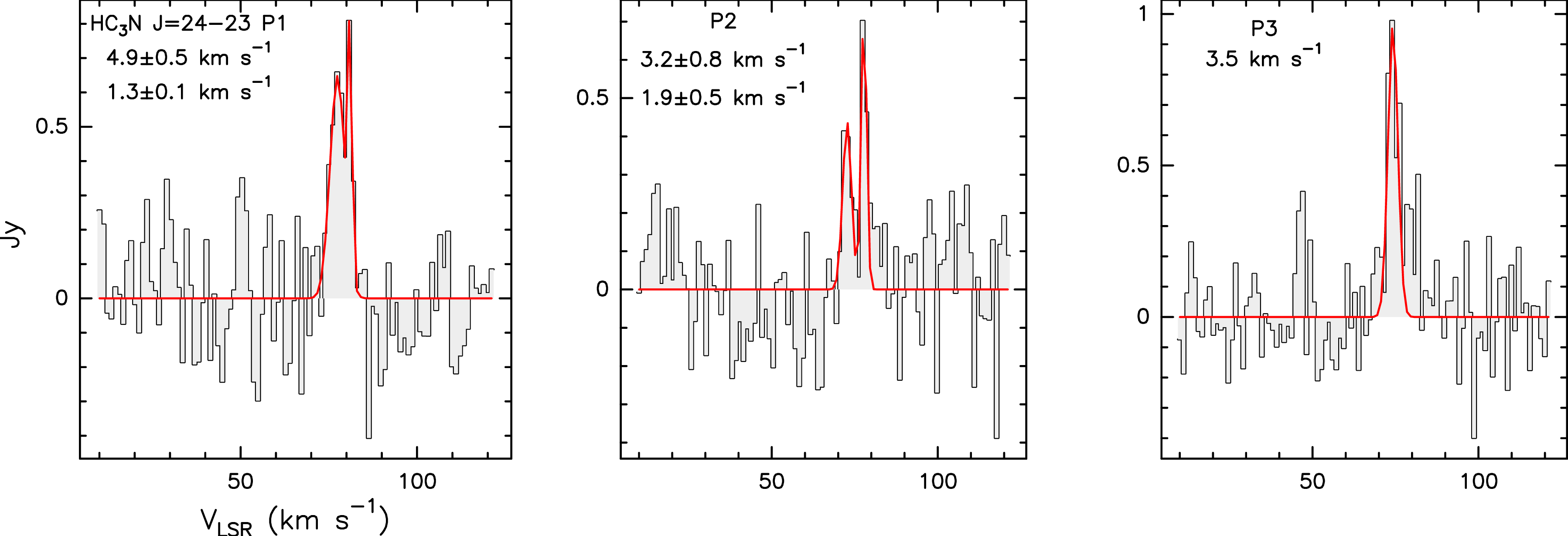}
\caption{Spectra extracted from positions P1, P2, and P3, indicated with white crosses in Fig.~\ref{fig-vel-70-85}, within the beam size of $\sim$4\asec.4$\times$4\asec.0 (black histrogram). The red solid lines are gaussian fits to the different velocity components for each region. Their best line widths are indicated in the top left corner of each panel.}
\label{fig-spectra-P1-P2-P3}
\end{figure*}

In this appendix we show the spectra extracted from single pixels (Fig.~\ref{fig-spectra-P1-P2-P3}), i.e. from the beam size of $\sim$4\asec.4$\times$4\asec.0, in positions P1, P2, and P3, indicated as white crosses in Fig.~\ref{fig-vel-70-85}. These regions have been chosen randomly near C2 and C3, where the intensity is not strong, to try to avoid possible contamination of different gas components of the condensations. The single pixel emission is even narrower (FWHM$\sim$1.3--4.9 km s$^{-1}$) than from the entire condensations C2 and C3 (FWHM$\sim$7.2~km~s$^{-1}$ and 8.8~km~s$^{-1}$). We discuss this result and its possible implications in Sect.~\ref{sec-evolstage}.

\clearpage
\section{Generation of \texttt{RADEX} synthetic spectra and implementation into \texttt{MADCUBA}}
\label{sec:app-radex}

To produce non-LTE synthetic spectra, we first run \texttt{RADEX} \citep{vandertak2007}, which used the data included in the Leiden Atomic and Molecular Database (LAMDA; \citealt{vandertak2007}){\footnote{Leiden Atomic and Molecular Database (LAMDA); \url{https://home.strw.leidenuniv.nl/~moldata/}}}. We assumed in all cases a static spherical geometry (using the expressions for the escape probability for the uniform sphere), which for HC$_3$N and the optically thin limit provides similar results to other geometries (e.g., expanding sphere). We used H$_2$ as collisional partner, using the collision coefficients by \citet{faure2016}.

We run the offline version of \texttt{RADEX} using a python-based script (presented in Sect. \ref{python-script}) to generate grids of physical parameters, $T_{\rm kin}$ and $n_{\rm{H}_2}$ (see Table \ref{table-grid}). 
We then used a newly implemented \texttt{MADCUBA} tool called \texttt{Import SPECTRA RADEX FILE}, which reads the output \texttt{RADEX} files (.out) to generate a synthetic spectrum to be compared with the observed spectrum. The tool allows to select the velocity at which the transitions will be centered, and the velocity/frequency sampling.
From the line intensities derived by \texttt{RADEX}, \texttt{MADCUBA} generates Gaussian line profiles centred at the chosen velocity with the line width used in the \texttt{RADEX} simulations, and properly considering the line opacity. The output from \texttt{RADEX} is imported into \texttt{MADCUBA} in units of brightness temperature, and the intensities are convolved in the same way that \texttt{SLIM} does for the simulated spectra as explained in Sect.~\ref{res-LTE}. The spectra are generated with the spectral axis in frequency like the observed data. 
For each transition of the molecule, a synthetic spectrum is produced and stored in the \texttt{MADCUBA} SPECTRA CONTAINER (see \citealt{martin2019}) to be compared with observations.

\subsection{Phyton-based script to generate grids of physical parameters with \texttt{RADEX}}\label{python-script}
\label{python-script}

The script below creates \texttt{RADEX} input ".inp" files from a grid of parameters to be used to subsequently run a series of \texttt{RADEX} models. The information needed for the script are: the name of the molecule to search in the \texttt{RADEX} collision coefficient folder ("molec"), the grid of densities and temperatures ("densities" and "tkin"), and the parameters to keep constant in the routine (background temperature "tbg", column density "cdmol", line width "dv", and the range of frequency to simulate the spectra between "fmin" and "fmax"). The script will finally return output ".out" files that can be read with \texttt{MADCUBA} to show the simulated spectra.
Below we give an example of the script used to simulate the spectra for the broad component C1.
\onecolumn

\begin{lstlisting}[language=Python]
import math
import os

molec = "hc3n" 

densities = [1e4, 2e4, 3e4, 4e4, 5e4, 6e4, 7e4] 
tkin=[100, 110, 120, 130, 140, 150, 160]
tbg = 2.73  
cdmol = 6.6e14  
dv = 23.6  
fmin = 2  
fmax = 336  


for i in range(len(densities)):
    for j in range(len(tkin)):

        file = molec + str(densities[i]) + str(tkin[j]) + '.inp' 

        with open(file, 'w') as f_output:
            f_output.write(molec + '.dat\n')
            f_output.write(molec + str(densities[i]) + str(tkin[j]) + '.out\n')
            f_output.write(str(fmin) + ' ' + str(fmax) + '\n')
            f_output.write(str(tkin[j]) + '\n')
            f_output.write('1\n')
            f_output.write('H2\n')
            f_output.write(str(densities[i]) + '\n')
            f_output.write(str(tbg) + '\n')
            f_output.write(str(cdmol) + '\n')
            f_output.write(str(dv) + '\n')
            f_output.write('0\n')
     
        os.system('radex <' + file)


\end{lstlisting}

\end{appendix}
\end{document}